%% file: Turner_estimating_tau_scatt.tex
\documentclass[twocolumn, times]{aastex63}

\shorttitle{Scattering Delay Mitigation: Cyclic Spectroscopy Techniques}
\shortauthors{Turner, J.\ E., {\em et al.} }

\usepackage{amsmath}
\usepackage{amssymb}
\usepackage[caption=false]{subfig}
\usepackage{xcolor}

\def\simless{\mathbin{\lower 3pt\hbox
   {$\rlap{\raise 5pt\hbox{$\char'074$}}\mathchar"7218$}}} 
\def\simgreat{\mathbin{\lower 3pt\hbox
   {$\rlap{\raise 5pt\hbox{$\char'076$}}\mathchar"7218$}}} 

\def\be{\begin{equation}}
\def\ee{\end{equation}}
\newcommand{\bea}{\begin{eqnarray}}
\newcommand{\eea}{\end{eqnarray}}

\begin{document}

\title{Scattering Delay Mitigation in High Accuracy Pulsar Timing: Cyclic Spectroscopy Techniques}









\input{authors}

\input{abstract}

\section{Introduction}
\par High-accuracy pulsar timing has been a transformative technique across a wide range of astrophysical fields, including neutron star mass measurements, binary star evolution, exacting tests of general relativity, pulsar astrometry, and  studies of the interstellar medium (ISM).
Now, in the era of pulsar timing arrays (PTAs), astronomers are poised to explore a gravitational-wave background due to supermassive black hole binaries (SMBHBs) located in galaxies at cosmic distances. Hints at the existence of such a background are already emerging in the data sets of PTAs through the presence of a common red noise process in the times-of-arrival (TOAs) of pulsars observed by the three major world-wide PTA collaborations \citep{nano_12,epta_back,ppta_back, 2022MNRAS.510.4873A}.
Such studies require attention to a myriad of details and careful understanding and correction for systematic effects due to a wide variety of sources: Earth rotation irregularities, solar system ephemeris inaccuracies, and even atomic time wander relative to an ensemble of highly accurate pulsar clocks \citep{nanotime}.
Propagation of  radio waves from pulsars to the Earth through the ionized, inhomogeneous ISM is a substantial source of noise, if not modeled properly, because the line of sight (LOS) from pulsar to Earth moves with respect to the medium due to motion of the endpoints and, subdominantly, motion of the medium itself \citep{nanotime,jones_dm,levin_scat,turner_scat}.
\par The major contributor to ISM-induced timing delays is  due to frequency-dependent ($\nu^{-2}$, where $\nu$ is the observing frequency) cold plasma dispersion along the LOS. This phenomenon has been studied in great detail since the early days of pulsar timing and can largely be corrected for, although important subtleties remain (e.g.\ \citet{css16}). However, multi-path propagation through the inhomogeneous ISM, or scattering, results in time-variable perturbations to pulsar TOAs. The resulting delays are expected to be proportional to $\nu^{-4.4}$ for a homogeneous Kolmogorov medium (although power laws ranging from around $-2.5$ to around $-4.5$ have been reported \citep{bcc+04,Bansal_2019,levin_scat,turner_scat}), and can be discerned via the delay of and structural broadening in an observed pulse. Effects of scattering, although understood theoretically and observed empirically in many high-accuracy timing programs, are not generally mitigated in major timing programs such as the NANOGrav (North American Nanohertz Observatory for Gravitational Waves) PTA and other global PTA efforts. As PTAs make their first detections and begin characterizing the low-frequency gravitational wave sky, it will be important to mitigate all possible delays. The goal of this paper and  subsequent work that we envision over the next several years is to develop effective mitigation strategies for time-variable scattering delays.
\par Cyclic spectroscopy (CS; \citet{dem11}) is central to our approach to this problem. CS is a powerful signal-processing technique that is already well-known and frequently used in the engineering community \citep{cyc1,cyc2,cyc3,ANTONI2007597} and applicable to periodic signals such as those from pulsars. In the few studies since its introduction to pulsar timing by \cite{dem11}, CS has been successful at producing high-resolution pulsar secondary spectra \citep{Walker_2013}, scattering measurements using CS-enabled fine channelization \citep{Archibald_2014}, and simulated recovery of the impulse response function (IRF) corresponding to a pulsar signal's passage through the ionized ISM \citep{Palliyaguru_2015}.
As detailed in \citet{dsj+20}, using CS to fully recover the IRF of the ISM, although a good long-term goal, has requirements, particularly signal to noise ratio (S/N), that are often not met with the current generation of radio telescopes. Here, we present a CS-derived quantity, $\tau_{\textrm{CS}}$, obtained from CS-based recovery of the IRF, which is more highly correlated with total scattering delay than other commonly utilized estimators. This work serves as proof of concept for the recoverability of scattering-based delays with CS, sometimes in conjunction with an autocorrelation function (ACF) estimator, addressing concerns about the accuracy of ACF-based estimators raised by authors such as \cite{coles_2010}.
\par The organization of this paper proceeds to a presentation of the basic theoretical framework in \S\ref{sec:theory}. Following this, we present the methodology and the results  of a simulation in which we compare the effectiveness of $\tau_{{\rm CS}}$ to other estimators of scattering delay, specifically the widely used estimators based on the ACF of the scintillated spectrum in \S\ref{sim_meth} and \S\ref{sec:sim}, respectively. We conclude with a discussion of future possibilities in \S\ref{sec:future}.
\section{Theoretical Basics}
\label{sec:theory}
As is standard practice in pulsar studies, we adopt an amplitude modulated noise (AMN, \citet{ric75}) model for the pulsar signal.
The electric field (single polarization) can then be represented as
\begin{equation}
\label{e_field}
E(t) = [p(t)n(t)] * h(t) + n_{\textrm{sys}}(t),
\end{equation}
\noindent where $p(t)$ is the original pulse profile at time $t$ mod $P$, with $P$ being the pulse period, $n(t)$ is the intrinsic modulated pulsar noise, $h(t)$ is the IRF, $n_\textrm{sys}(t)$ is the noise from the sky and receiver present in the system, uncorrelated across pulse periods, and we have used the notation of \cite{dsj+20}. We choose to represent the signal as complex-valued, hence $N(t)$, $h(t)$, and $n_{\textrm{sys}}(t)$ are complex. Additionally, we can write this electric field as
\begin{equation}
E(t) = E_0(t) * h(t) + n_{\textrm{sys}}(t),
\end{equation}
\noindent where which $E_0(t) = p(t)n(t)$. The corresponding frequency domain signal model is
\begin{eqnarray}
E(\nu) & = & [p(\nu) * N(\nu)]H(\nu) + N_{\textrm{sys}}(\nu) \\
& = & E_0(\nu)H(\nu) + N_{\textrm{sys}}(\nu),
\end{eqnarray}

\noindent where we use $E_0(\nu)$ instead of $p(\nu) * N(\nu)$ because the convolution occurs upon emission at the pulsar. $H(\nu)$, which is the Fourier transform of $h(t)$, is the transfer function (TF) of the ISM.

\par The resulting cyclic spectrum of $E(t)$ is
\begin{equation}
\label{cyc_eq}
{S_E}(\nu, \alpha_k) = \langle E(\nu + \alpha_k/2)E^{*}(\nu - \alpha_k/2)\rangle,
\end{equation}
\noindent where $\nu$ is the bandpass frequency at which the signal is measured and $\alpha_k = k/P$ is the cyclic frequency, also known as the modulation frequency, and the average is over an integer number of pulses. The cyclic spectrum is a complex-valued function with amplitude and phase for each $(\nu, \alpha_k)$ pair, and is undefined for non-periodic signals. It is important to keep in mind that, in practice, Equation \ref{cyc_eq} is averaged over a period of time over which the transfer function must remain unchanged, which must be less than the diffractive timescale of the ISM along a particular LOS.  

\par If we make the assumption that a scattering delay can be seen in a pulse profile as a translation in the time domain, then as a consequence of the shift theorem of Fourier Transforms this results in a phase slope in the frequency domain. For this reason, it can be useful to examine the CS phase slope, $\phi_{\textrm{cyc}}(\nu, \alpha_k)$, which is found via
\begin{equation}
\label{cyc_phase}
\phi_{\textrm{cyc}}(\nu, \alpha_k) = \tan^{-1}\Bigg(\frac{{\rm Im}\{S_E(\nu, \alpha_k)\}}{{\rm Re}\{{S_E}(\nu, \alpha_k)\}}\Bigg).
\end{equation}
It can also be found by examining $\phi_H$, the phase of the transfer function

\begin{align}
\begin{split}
\label{phase_oth}
\phi_{\textrm{cyc}}(\nu, \alpha_k) &= \phi_H(\nu + \alpha_k/2)-\phi_H(\nu - \alpha_k/2)\\  &= \alpha_k \frac{\phi_H(\nu + \alpha_k/2)-\phi_H(\nu - \alpha_k/2)}{\alpha_k}.
\end{split}
\end{align}

Under the assumption that the cyclic frequency $\alpha_k$ is much less than the diffractive bandwidth, $\Delta \nu_{\textrm{d}}$, we can make the approximation 

\begin{equation}
\label{phase_approx}
\phi_{\textrm{cyc}}(\nu, \alpha_k) \approx \alpha_k \frac{d\phi_H}{d\nu} = \frac{k}{P}\frac{d\phi_H}{d\nu}.
\end{equation}

When the S/N is large enough, the transfer function phase can be recovered by simply integrating the CS phase, 
\begin{equation}
\label{phase_int}
\phi_H(\nu, \alpha_k) = \int \frac{\phi_{\textrm{cyc}}(\nu, \alpha_k)}{\alpha_k}d\nu.
\end{equation} 
At lower S/N ratios this is not possible and more sophisticated recovery algorithms are required \citep{dem11,Walker_2013}. The transfer function amplitude for a given $\alpha_k$ can then approximated as the square root of the CS amplitude for that $\alpha_k$. Finally, the reconstructed transfer function can then be inverse Fourier transformed back into a reconstructed IRF, and the recovered scattering delay can be found by calculating the centroid of the intensity IRF,

\begin{equation}
\label{recovered_tau}
\tau_{\textrm{CS}}(\alpha_k) = \frac{\int t \ |h_{\textrm{CS}}(t, \alpha_k)|^2dt}{\int |h_{\textrm{CS}}(t,\alpha_k)|^2dt}.
\end{equation}

\section{Simulation Methodology}
\label{sim_meth}
Our simulations began by creating real and imaginary components of white noise, which we call $n(t)$, and multiplying them by a complex, one-sided decaying exponential to form our IRF,
\begin{equation}
\label{h(t)}
h(t)= \textrm{Re}\{n(t)\}e^{-t/2\tau}+ \textrm{Im}\{n(t)\}ie^{-t/2\tau},
\end{equation}
where the length $t$ is the value in time at which the one-sided exponential is sampled. The inclusion of this amplitude-modulated white noise, which varies from realization to realization, serves to mimic how the ISM changes over the course of many observations by emulating the time-varying effects of scintillation \citep{Narayan}. Each realization corresponds to a pattern of scintles, meaning $h(t)$ can be considered constant on scales shorter than the scintillation time. 
The injected value of the scattering delay for a given realization is given by the centroid of the resulting pulsar signal, $\tau_{\textrm{cent}}$, which can be found via
\begin{equation}
\label{toa_cent}
\tau_{\textrm{cent}} = \frac{\int t \ |h(t)|^2dt}{\int |h(t)|^2dt}.
\end{equation}
It is worth noting that, in real observations, scattering variations are in fact correlated with each other, since scintillation is often dominated by compact structures at a fixed or close-to-fixed angular position that moves between observations, primarily as a consequence of a pulsar's proper motion \citep{hsa+05}.
\par For simplicity we treat the pulse profile $p(t)$ as a delta function of height unity. This model acts as a best-case scenario pulse, removing all other factors that might interfere with our aim of solely comparing the effectiveness of various estimators at recovering a delay imparted by the ISM on radio signals of varying strengths. We appreciate that this technique would not be necessary for a true delta-like pulse, as if $p(t)$ is truly delta-like, then the IRF can also be obtained directly, as was shown using narrow pulses from PSR B1957+20, where consistent delay values have been obtained from fitting a recovered IRF and from scintillation bandwidth measurements \citep{Main_2017}.

\begin{figure}[!ht]
\includegraphics[scale=.58]{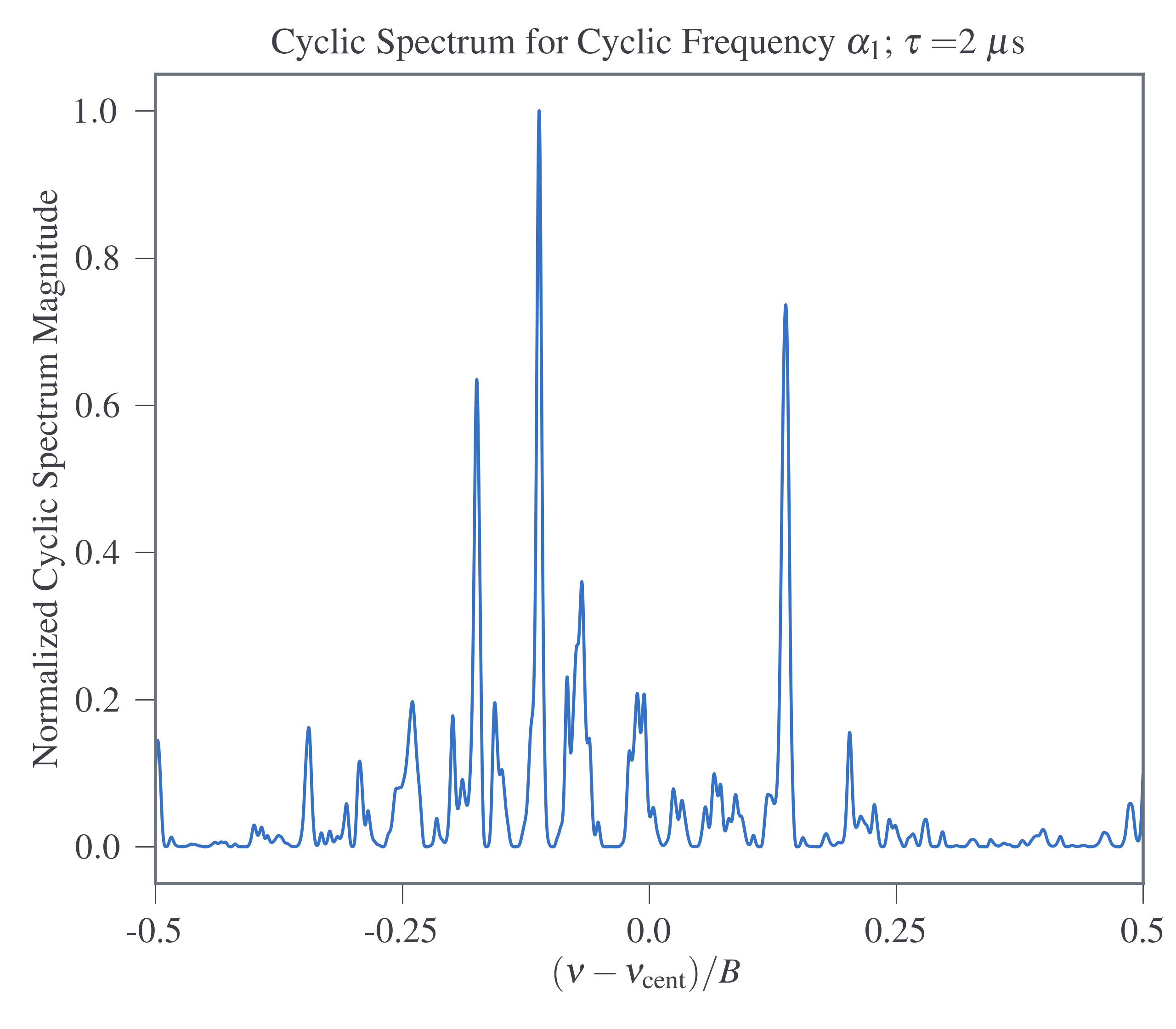}
\centering
\caption{An example normalized cyclic spectrum as a function of the normalized bandpass taken at the cyclic frequency $\alpha_1$ for a simulated scattering delay of 2 $\mu$s using a spin period of 2 ms and a sampling interval of 100 ns. Here $\nu_{\textrm{cent}}$ is the center frequency of the observation and $B$ is the observing bandwidth.}
\label{cyc_plot}
\end{figure}

\begin{figure}[!ht]
\includegraphics[scale=.52]{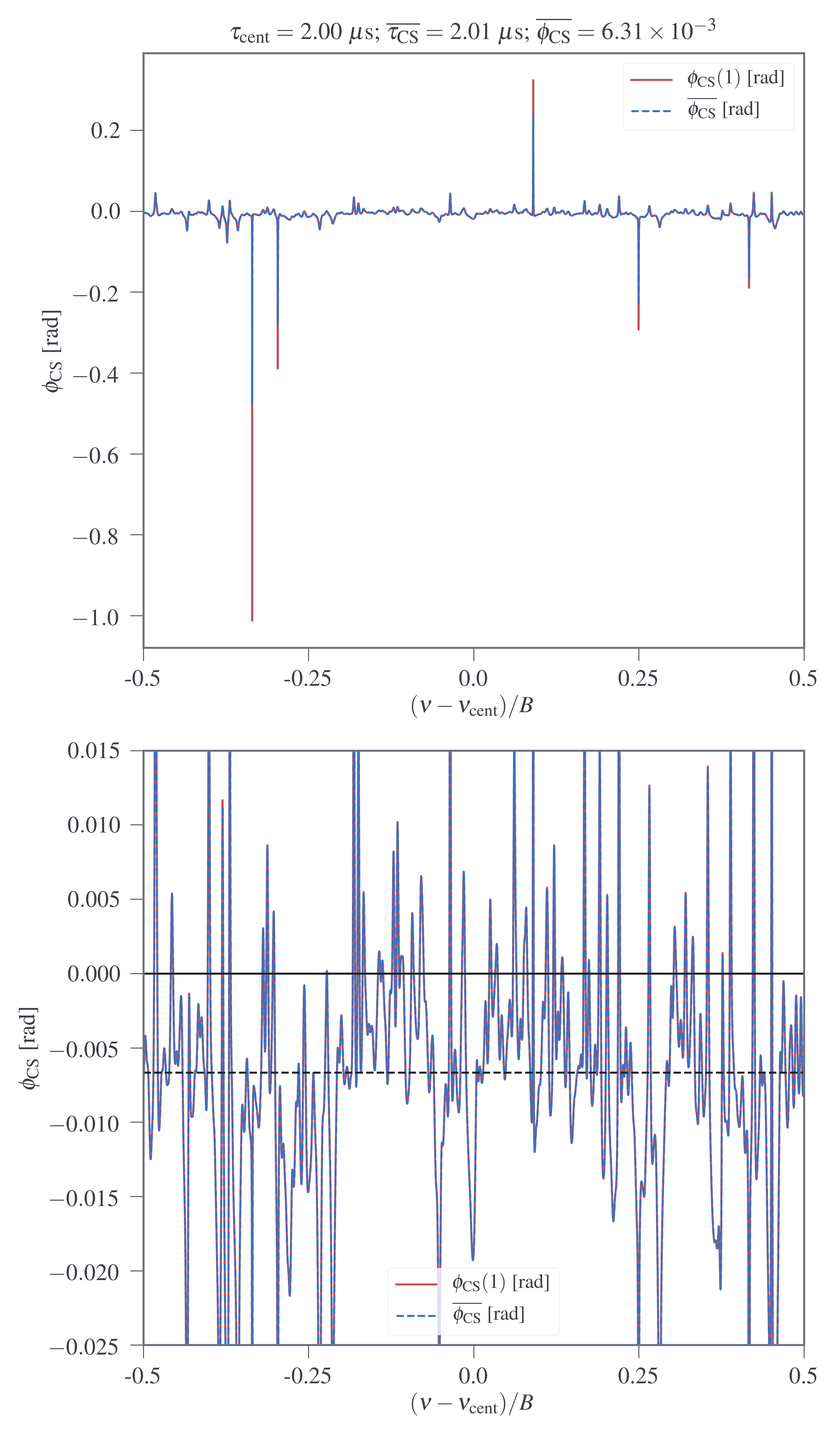}
\centering
\caption{(Top) An example cyclic phase as a function of the normalized bandpass taken at the cyclic frequency $\alpha_1$ (red) as well as using the weighted average of the first 50 cyclic frequencies (dashed blue) for a simulated scattering delay of 2 $\mu$s using a spin period of 2 ms and a sampling interval of 100 ns, corresponding to $P=2$ ms. (Bottom) A zoomed in version of the top plot to better visualize structure. The dashed black line indicates the average cyclic spectrum phase, while the solid black line indicates a phase of zero. As seen in the top figure, the phase only utilizing the first cyclic frequency has much more extreme outliers. In fact, over many noise realizations at a S/N of 10, weighted average cyclic phases using 50 cyclic frequencies typically exhibit around 79\% smaller standard deviations compared to just the phase at the first cyclic frequency.}
\label{phase_plot}
\end{figure}

\begin{figure}[!ht]
\includegraphics[scale=.45]{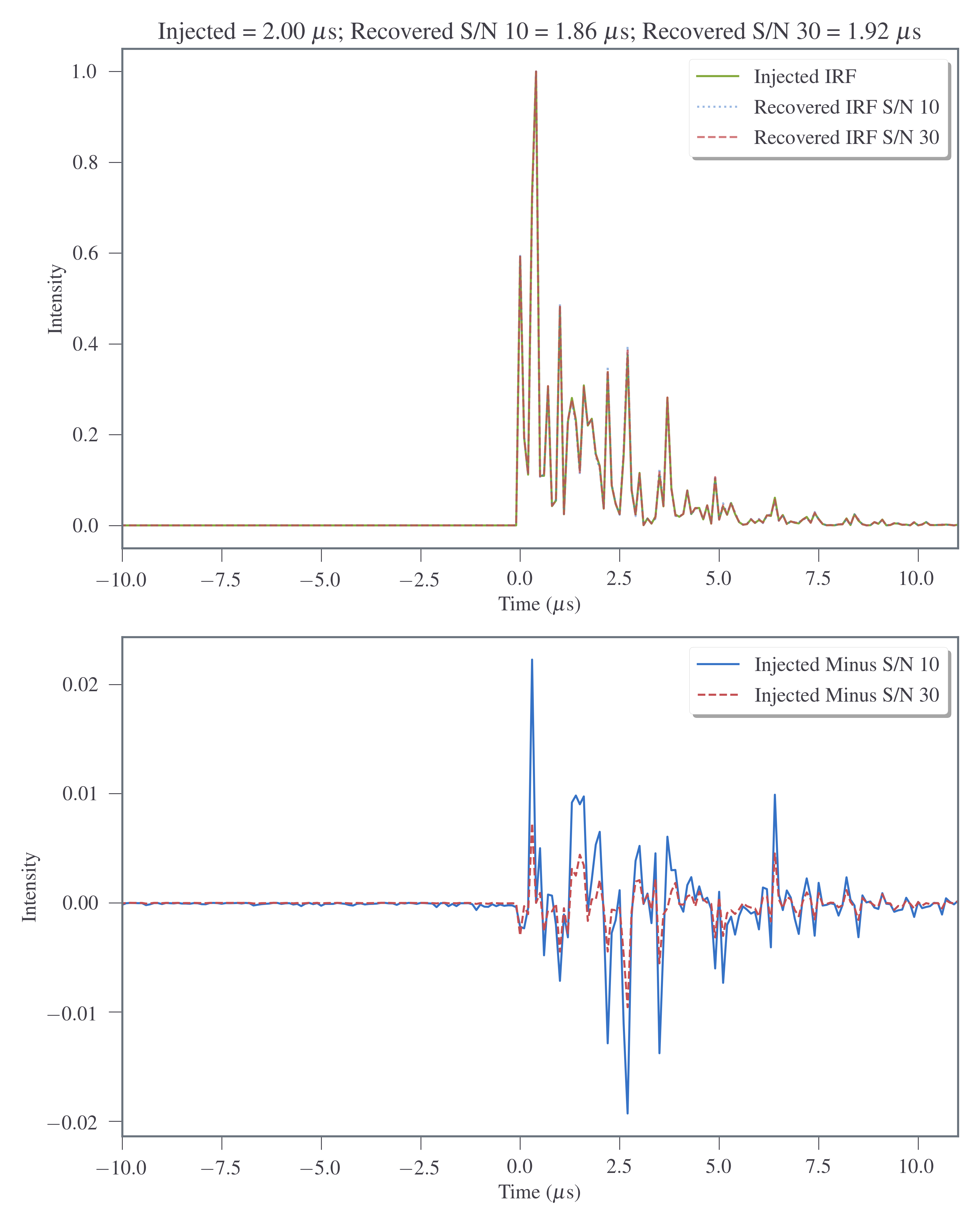}
\centering
\caption{(Top) An example injected IRF intensity prior to the inclusion of additive noise and the corresponding recovered IRF intensities using information from only the first cyclic frequency using a S/N of 10 and 30. (Bottom) Differences between the noiseless injected IRF intensity and recovered IRF intensities at a S/N of 10 and 30. Despite recovering the IRF quite well at these S/N, small differences between the injected and recovered IRF intensities can lead to noticeable differences in the recovered delay.}
\label{irf_compare}
\end{figure}

We then follow Equations \ref{e_field}-\ref{cyc_eq}, deviating only in that our noise is added only once we are in the frequency domain, to get the cyclic spectrum, an example of which can be seen Figure \ref{cyc_plot}. Next, the cyclic spectrum phase is calculated using Equation \ref{cyc_phase}. An example cyclic phase plot can be seen in Figure \ref{phase_plot}. We then calculate $\tau_{{\rm CS}}$ by following the methodology described after Equation \ref{phase_oth} and up to Equation \ref{recovered_tau}. An example injected and recovered IRF intensity can be seen in Figure \ref{irf_compare}. It is important to note that, in our simulations, $p(t)$ and $h(t)$ are identical for each pulse, while $n(t)$ and $n_{\textrm{sys}}(t)$ are randomized. If there were significant variations of $p(t)$ beyond AMN, we suspect our method would still be accurate, although CS in general becomes decreasingly effective as $p(t)$ gets wider and S/N gets lower, which we believe are more significant factors.

\par In real pulsar data, the Fourier coefficients, $A_k$, of a pulse drop off at higher harmonics, with a non-scattered CS effectively being the Fourier transform of the pulse shape and more or less constant in radio frequency. For this reason, we weight our transfer function at the $k^{\textrm{th}}$ cyclic frequency by the corresponding $k^{\textrm{th}}$ Fourier coefficient of a pulse with a reasonable period and width. In this simulation we chose a period of 2 ms and a width of 110 $\mu$s. This pulse width was chosen simply because it is the pulse width of PSR J1713+0747 \citep{parkes}, which has a sharp pulse that can be well-approximated as a Gaussian. Effectively, our simulation is using the Fourier coefficients of a slightly faster rotating PSR J1713+0747. 
\par The precision of the recovered delay estimation improves as we utilize more delays from higher cyclic frequencies, although the number of cyclic frequencies that have usable information depends on a number of factors, including the S/N of the pulsar signal and the pulsar duty cycle. For these simulations we make use of the first 50 cyclic frequencies in the cyclic spectra and, to calculate $\tau_{{\rm CS}}$ for a given noise realization, take a weighted average of the recovered delays from these cyclic frequencies, with the weight at the $k^{\textrm{th}}$ cyclic frequency being the $k^{\textrm{th}}$ $A_k$ value of the pulsar signal mentioned above.

\par We then compared this estimator to the more traditional methods of recovering scattering delays, which involve calculating the ACF of a dynamic spectrum, or the intensity of the pulsar signal in both frequency and time. The changes in the intensity of the dynamic spectrum over frequency and time can create patchy features known as scintles, and the corresponding scattering delay associated with a given scintle is inversely proportional to that scintle's width in frequency. An ACF is able to pick up on a dynamic spectrum's scintillation pattern, with the width of the ACF's central peak then being relatable to the typical scintle width in that dynamic spectrum. The dynamic spectrum in the case of our delta-function pulse with unity flux at all frequencies is simply $|E(\nu)|^2$, and has the form of Equation \ref{cyc_eq} for $\alpha=0$. From there, the ACF is found by normalizing the mean-subtracted filter function cross-correlated with itself,
\begin{align}
\label{acf_eq}
\textrm{ACF}(\nu) &= \Bigg[\sum_{\nu} \Bigg(|E(\nu)|^2-\overline{|E(\nu)|^2}\Bigg) \nonumber \\ &\times \Bigg(|E(\nu+\Delta \nu)|^2-\overline{|E(\nu+\Delta \nu)|^2}\Bigg)\Bigg] \nonumber \\ &\div\textrm{Max} \Bigg[\sum_{\nu} \Bigg(|E(\nu)|^2-\overline{|E(\nu)|^2}\Bigg) \nonumber \\ &\times \Bigg(|E(\nu+\Delta \nu)|^2-\overline{|E(\nu+\Delta \nu)|^2}\Bigg)\Bigg],
\end{align}
where the horizontal bar indicates we are taking the average.

\par
\begin{figure}[!ht]
\includegraphics[scale=.58]{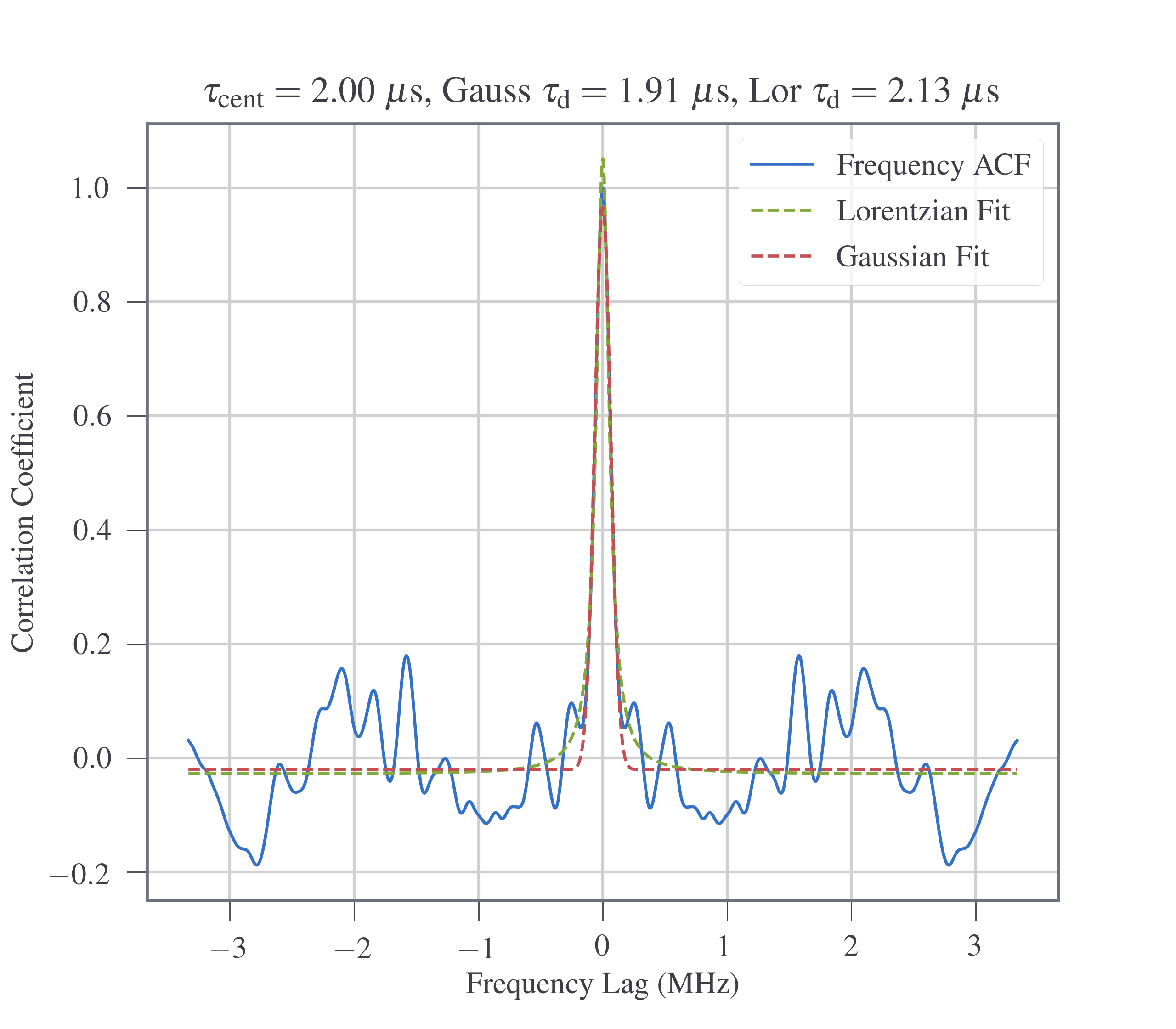}
\centering
\caption{The frequency ACF (blue) of a dynamic spectrum for a scattering delay of 2 $\mu$s. Green and red dotted lines correspond to fits to the ACF using Lorentzian and Gaussian distributions, respectively. Delays in the title correspond, from left to right, to injected $\tau_{\textrm{cent}}$ delay, the Gaussian ACF estimator, and the Lorentzian ACF estimator.}
\label{acf_ex}
\end{figure}

 The ACFs were then fit with both a Gaussian and Lorentzian distribution, and the scattering delay was found via 

\begin{equation}
\label{band_delay}
2\pi \Delta \nu_{\text{d}} \tau = C_1,
\end{equation}
where $\Delta \nu_{\textrm{d}}$ is the scintillation bandwidth, defined as the half-width at half-maximum of the ACF along the frequency axis, and ${C}_{{\rm{1}}}$ is a dimensionless quantity ranging from $0.6-1.5$ conditional on the geometry and spectrum of the electron density fluctuations of the medium \citep{cr98}. In this analysis we assume ${C}_{{\rm{1}}}=1$. Mathematically, using the Lorentzian to fit the ACF makes more sense because the Lorentzian distribution is the square of the Fourier transform of the one-sided decaying exponential \citep{cwb85}, although Gaussian distributions are close approximations that have been used in a number of scintillation studies \citep{Bhat_1999, wmj+05, levin_scat,turner_scat}. An example ACF fit with both Lorentzian and Gaussian distributions is shown in Figure \ref{acf_ex}.

\section{Simulation Results}
\label{sec:sim}
\begin{figure*}[!ht]
\includegraphics[scale=.58]{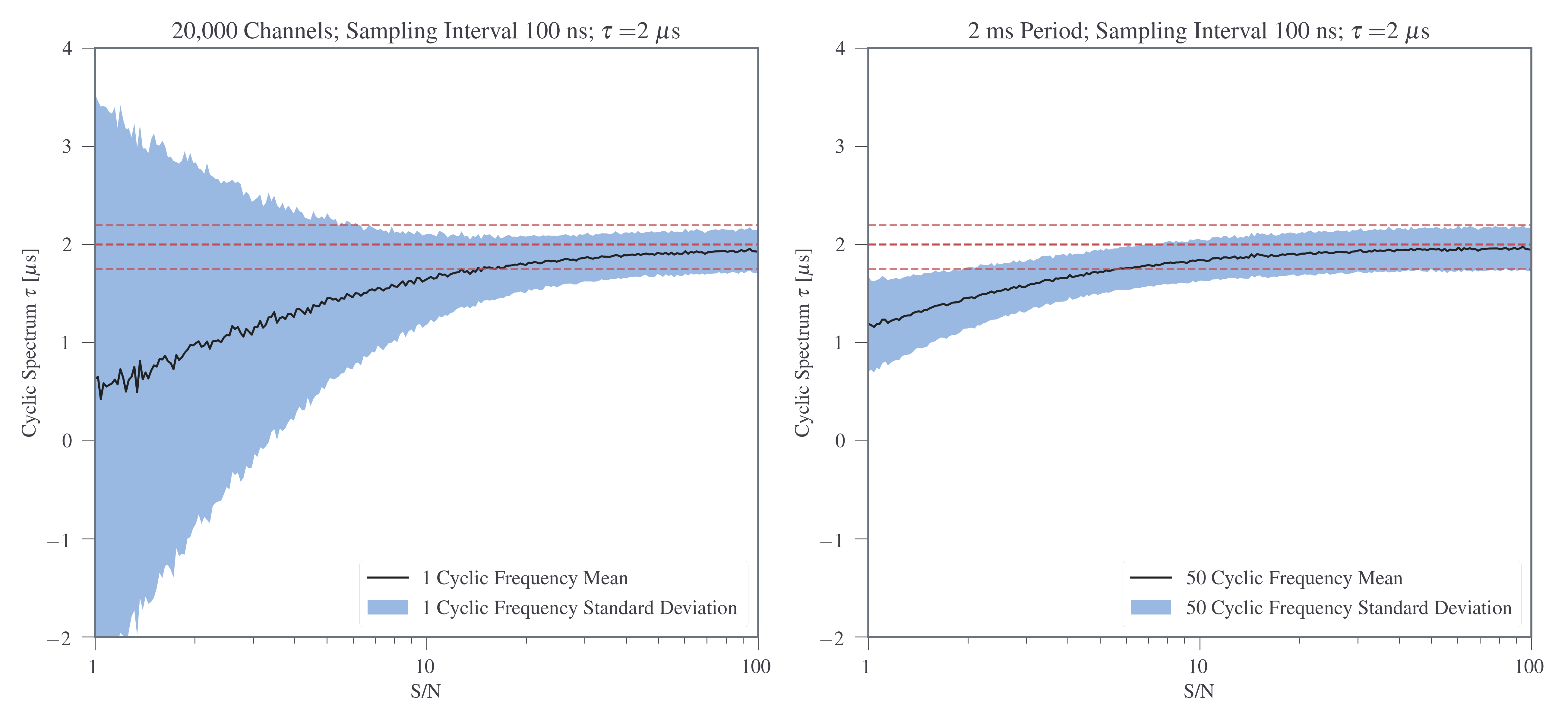}
\centering
\caption{Cyclic spectrum estimator simulation results for random noise draws for a $\tau$ of 2 $\mu$s for 300 values of S/N using a spin period of 2 ms and a sampling interval of 100 ns using one (left) and 50 (right) cyclic frequencies. The dashed red lines represent the mean plus or minus one standard deviation of the injected $\tau$ at each S/N, with the mean and standard deviation of the recovered values at each epoch shown in black and light blue, respectively.}
\label{cyc_scatt}
\end{figure*}

Our main simulation consisted of 1000 random noise draws for a $\tau$ of 2 $\mu$s using 20,000 time samples, $n_{\textrm{samp}}$, and a sampling interval, $s_{\textrm{int}}$, of 100 ns, corresponding to $P=2$ ms. Here, $n_{\textrm{samp}}$ refers to phase bins rather than baseband voltage samples. This framework assumes we are using baseband data recording with a bandwidth of 10 MHz. For larger bandwidths on the order of hundreds of MHz, individual scintles get progressively wider at higher frequencies, which other studies have compensated for by ``stretching" the entire dynamic spectrum \citep{levin_scat,turner_scat}. In these studies, the spectrum is scaled by $\nu^{-\beta}$, with $\beta$ being the scattering scaling index determined for a given pulsar's LOS, relative to the center frequency to give all scintles approximately equal width across the band. This small 10 MHz bandwidth was chosen to avoid the scintle stretching that would be required at larger bandwidths. Additionally, this bandwidth and scattering delay combination results in a similar number of scintles on average across the band as is seen for many NANOGrav pulsars \citep{turner_scat}, meaning that we have approximately the same amount of data informing our ACFs, and consequently similar precision for a comparable S/N. This series of 1000 random noise draws was repeated over 300 different values of S/N ranging from around 0.3 to 100, with the S/N defined as the square of the inverse of the standard deviation of $N_{\textrm{sys}}(t)$, since our transfer functions are normalized prior to noise being added.

\begin{figure*}[!ht]
\includegraphics[scale=.58]{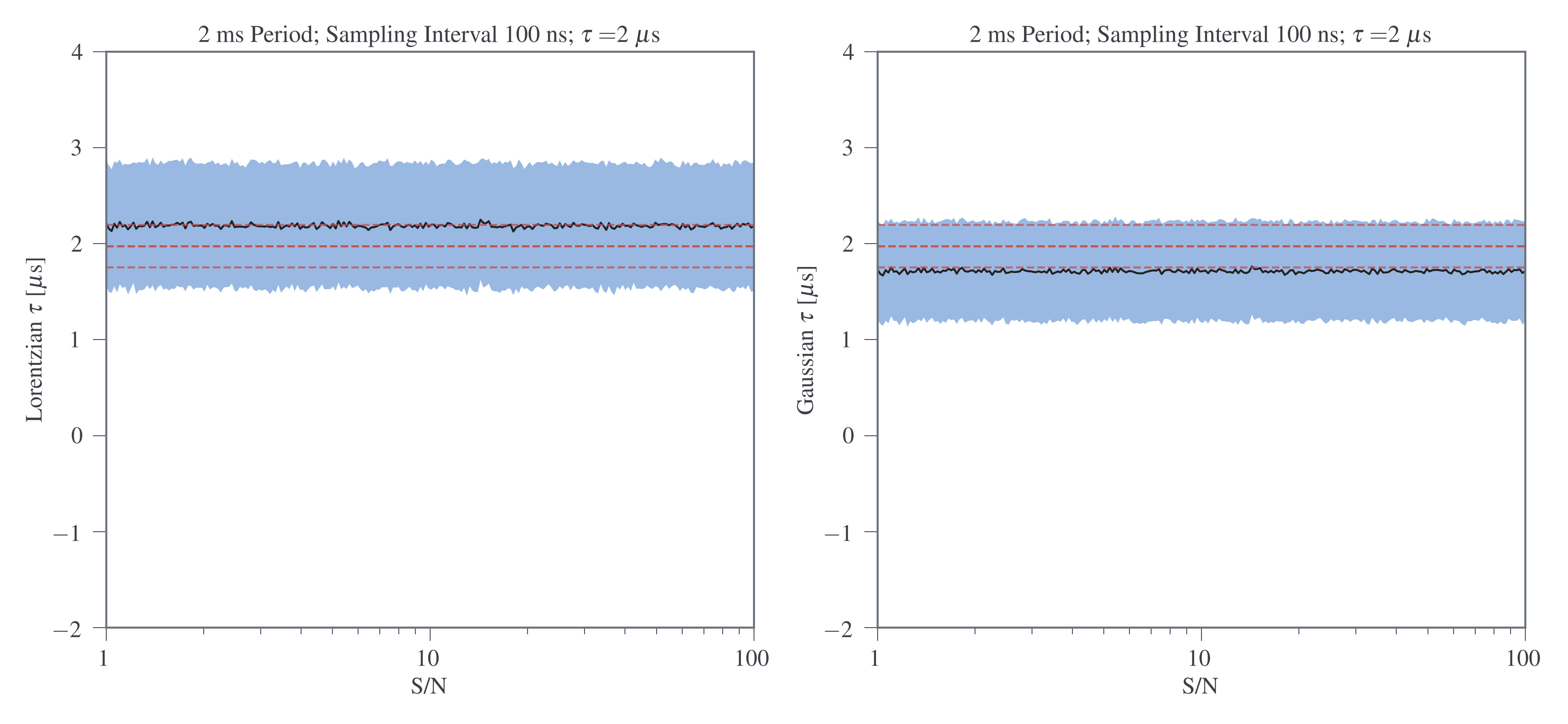}
\centering
\caption{Lorentzian (left) and Gaussian (right) ACF estimator simulation results for random noise draws for a $\tau$ of 2 $\mu$s for 300 values of S/N using a spin period of 2 ms and a sampling interval of 100 ns. The dashed red lines represent the mean plus or minus one standard deviation of the injected $\tau$ at each S/N, with the mean and standard deviation of the recovered values at each epoch shown in black and light blue, respectively.}
\label{acf_scatt}
\end{figure*}

\par The results of these simulations using the cyclic spectrum and Lorentzian and Gaussian ACF estimators for the recovery of $\tau$ are shown in Figures \ref{cyc_scatt}, \ref{acf_scatt}, respectively. The cyclic spectrum estimator using 50 cyclic frequencies appears to converge to a stably recovered value of $\tau$ at a S/N around 100, while the two ACF estimators have already converged at the lowest S/N in our simulation, which may imply that, given sufficient frequency resolution, there appears to be a range of lower S/N where these estimators are superior to the cyclic spectrum estimator. 
\par In fact, the lack of improvement in the ACF estimators demonstrates that good frequency resolution, specifically the ratio of the scintillation bandwidth to the overall observing bandwidth, and consequently, the total number of scintles across the observing band, is much more important than S/N for accurate ACF estimator recovery. This effect, known as the finite scintle error, can be determined via
\begin{equation}
\label{finite_scintle}
\begin{split}
\epsilon & \approx \tau_{\text{d}} N_{\rm{scint}}^{-1/2} \\
& \approx \tau_{\text{d}}[(1+\eta_{\text{t}}T/\Delta t_\text{d})(1+\eta_\nu B/\Delta \nu_\text{d})]^{-1/2},
\end{split}
\end{equation}
where ${N}_{{\rm{scint}}}$ is the number of scintles in the dynamic spectrum, $T$ and $B$ are total integration time and total bandwidth, respectively, $\Delta t_{\textrm{d}}$ is the scintillation timescale, defined as the half-width at $e^{-1}$ of the dynamic spectrum's ACF along the time axis, and ${\eta }_{{\rm{t}}}$ and ${\eta}_{\nu}$ are filling factors ranging from $0.1$ to $0.3$ depending on the definitions of characteristic timescale and scintillation bandwidth, and in our case both set to 0.2 \citep{cor86}. Since scattering delays only depend on the scintillation bandwidth, the scintillation timescale is not important for these simulations. As a result, for simplicity, we can assume these simulated observations had observing times much less than the scintillation timescale, and so Equation \ref{finite_scintle} can be reduced to
\begin{equation}
\label{red_fin_scint}
\epsilon \approx \tau_{\text{d}}(1+\eta_\nu B/\Delta \nu_\text{d})^{-1/2}.
\end{equation}
\par Taking the results of a typical 1000 sample run, we find that the average Lorentzian delay spread is around 0.66 $\mu$s and the average Gaussian delay spread is around 0.52 $\mu$s, while the average Lorentzian finite scintle error is around 0.40 $\mu$s and the average Gaussian finite scintle error is around 0.36 $\mu$s. Further, we did a series of tests in which we varied the sampling interval, hence the bandwidth, of the simulation. These tests verified that the spread of ACF values follows the $B^{-1/2}$ scaling of Equation \ref{red_fin_scint} in the many-scintle regime.
\par This limitation on the effectiveness of estimation ACF-based techniques also means that methods such as those demonstrated by \cite{hs08}, which estimate scattering delays by integrating along the differential delay axis of the secondary spectrum, will face the same constraints. Other non-CS techniques do exist to reconstruct the IRF, such as interstellar holography demonstrated by \cite{wksv08}, although this particular technique requires a high S/N. Our simulations have shown that an IRF recovery approach based on CS, albeit one that uses a simple, non-iterative phase reconstruction, is quite effective at moderate S/N. That being said, future work will be required to fully evaluate the relative merits of these different approaches.
\par Figures \ref{cyc_scatt} and \ref{acf_scatt} also show that all three estimators converge to roughly the correct value, although there are slight biases in the mean values for the ACF estimators, whereas none is seen for the cyclic spectrum estimator. Additionally, for an ideal estimator, at sufficiently high S/N the standard deviation in the recovered delays should end up matching the standard deviation in the injected $\tau_{\textrm{cent}}$ delays, which we see only in the cyclic spectrum estimator. For reasons that will be discussed later, we do not believe that the biases in the ACF estimators are simply an indicator that a different $C_1$ should be used for our choice of impulse response.
\par A significant difference is also noticeable in the mean and standard deviation at low S/N between using only one cyclic frequency and using 50 cyclic frequencies. While the single cyclic frequency estimator appears to converge at a similar, if not slightly higher, S/N, its standard deviation is still significantly larger than the 50 cyclic frequency estimator at lower S/N. Overall, this presents a strong argument that using many cyclic frequencies is superior. 
\par The extreme variability seen at low S/N in the one cyclic frequency cyclic spectrum estimator, and the trend toward average recovered delays around zero $\mu$s in both cyclic spectrum estimators, is the result of the white noise overwhelming an IRF that has both positive and negative components, resulting in a signal that is on average centered around zero on the time axis. As the IRF becomes more discernible from the additive white noise at higher S/N, a signal that is increasingly centered in a positive region on the time axis is recovered, resulting in positive recovered delay values. 

\par On a related note, if our ACF estimators did not have sufficient frequency resolution at lower S/N, the excess noise would have resulted in scintles appearing narrower and therefore yielding higher measured scattering delays, leading to the ACF estimators being biased high in addition to having large variability. This high bias is also a consequence of ACF fitting always producing a positive definite value, whereas the cyclic spectrum estimator's ability to return both positive and negative values results in more manageable behavior at low S/N. 

\begin{figure}[!ht]
\includegraphics[scale=.58]{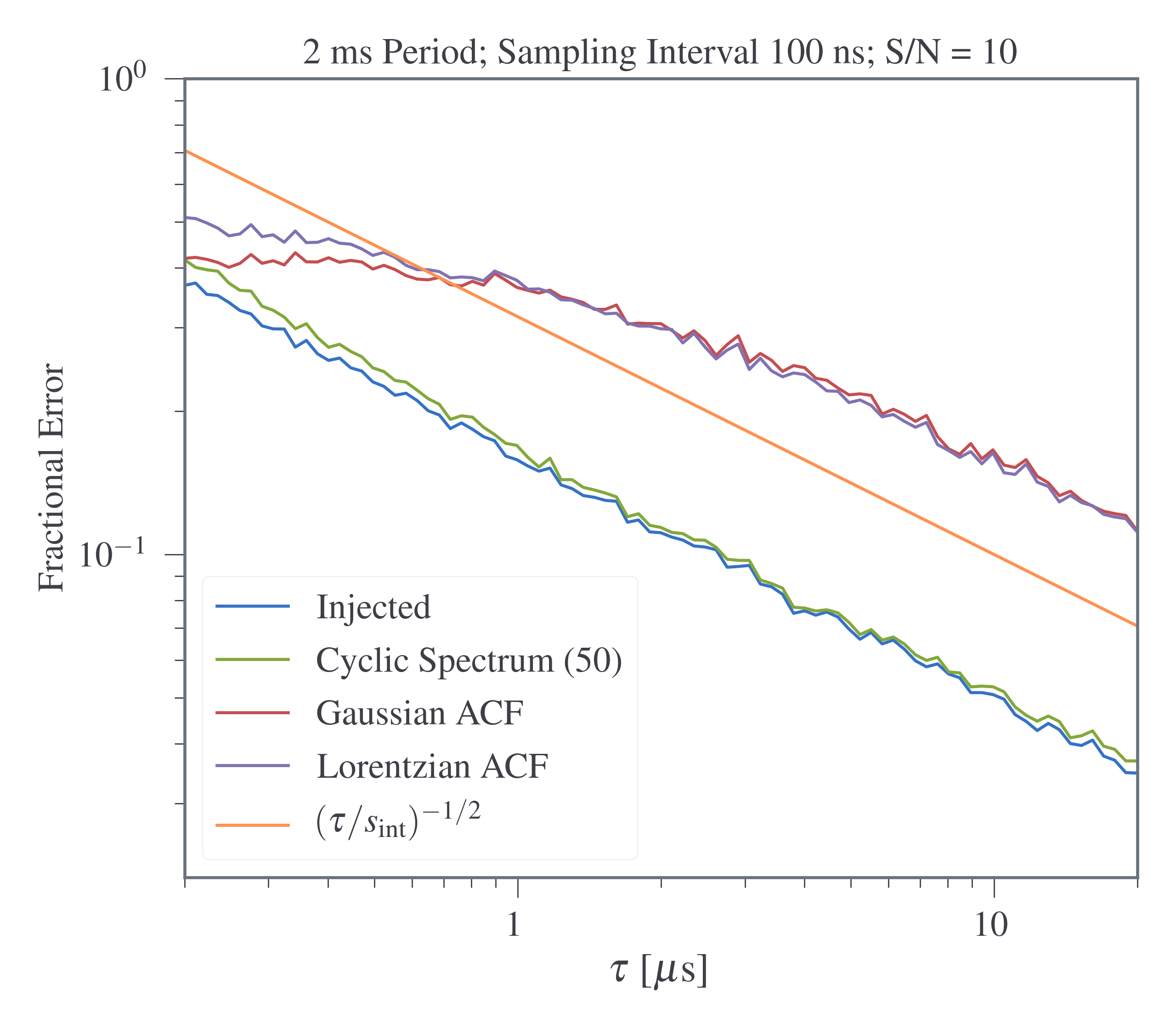}
\centering
\caption{The fractional error of the recovered delay for 100 values of $\tau$ with a sampling interval $s_{\textrm{int}}$ of 100 ns using a spin period of 2 ms with 50 cyclic frequencies at a S/N of 10. The fractional error scales as the inverse square root of the number of scintles across the observing band for a typical observation, or, equivalently, the inverse square root of the delay divided by the sampling interval. Throughout much of the abscissa range, the cyclic spectrum estimator has a fractional error approximately 64\% smaller than that of the ACF estimators. ACF estimators can be seen to flatten out in the smaller delay-to-sampling interval ratio regime as they are no longer able to detect a signal above the noise. The improvement in precision as the delay gets larger while maintaining this sampling interval demonstrates the benefits of proposed wider bandwidth observing programs.}
\label{samp_results}
\end{figure}
\par Supplemental simulations also show that, after reaching a sufficient S/N, additional gains in precision for all estimators are also partially limited by the ratio of the delay to the sampling interval, regardless of the number of time samples in use. This is under the assumption that we are already using a sufficient number of time samples such that accurate scattering estimations are possible. As shown in Figure \ref{samp_results}, when we run our simulation at a S/N of 10 at various values of delay with a constant sampling interval of 100 ns, we find a significant improvement in our fractional error (or in this case, the standard deviation of the recovered values divided by delay) as the delay-to-sampling interval ratio increases, following an inverse square root power law for all estimators. This quantity is also equivalent to the inverse square root of the number of scintles across the observing band for a typical observation. Since both the number of scintles across the observing band and the sampling interval are inherently tied to the maximum possible bandwidth we can utilize, i.e., the inverse of the sampling interval, these results provide strong support for the introduction of ultra wideband (UWB) observation programs.

\par
\begin{figure}[!ht]
\includegraphics[scale=.59]{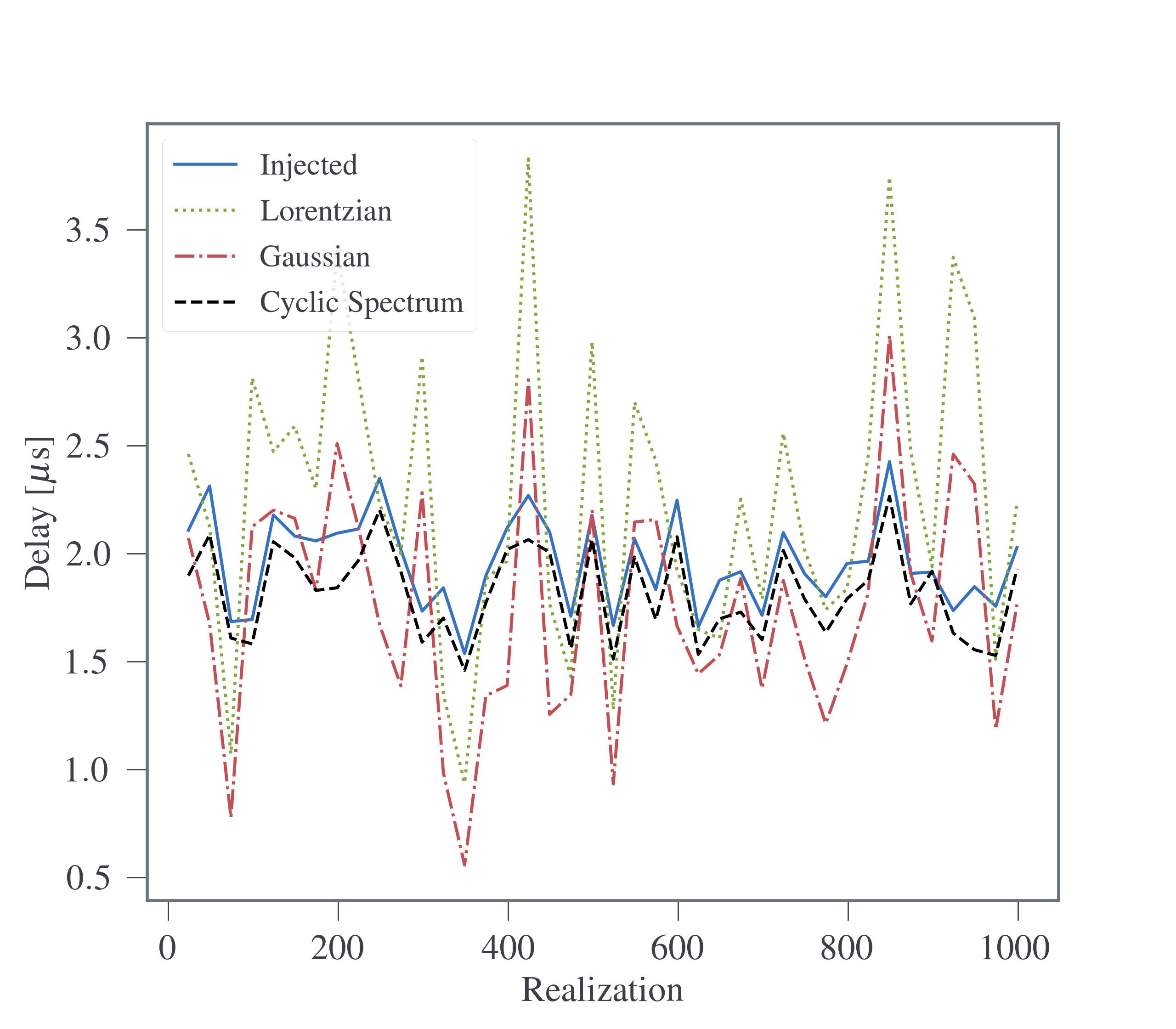}
\caption{A sample of realizations for a S/N of 10 for the simulation described above. The accuracy of the different estimators compared to the injected value found by $\tau_{\textrm{cent}}$ follows the behavior seen in Figures \ref{cyc_scatt}, and \ref{acf_scatt}.}
\label{coarse_samp}
\end{figure}

 In addition to examining the precision and accuracy over many realizations, we also looked at how this behavior tracked over individual realizations. A typical example of this at a S/N of 10 can seen in Figure \ref{coarse_samp}, where we show every 20th realization of the simulation for visual ease. While all three estimators generally follow the injected delays, the cyclic spectrum estimator clearly tracks these injected values much better than the ACF estimators.



\begin{figure*}[!ht]
\centering
\subfloat[Cyclic spectrum estimator using 50 cyclic frequencies. The slight offset from the line of equality is expected at a S/N of 10, as our 50 cyclic frequency CS results in Figure \ref{cyc_scatt} were not fully converged at this S/N.]{\includegraphics[scale=.54]{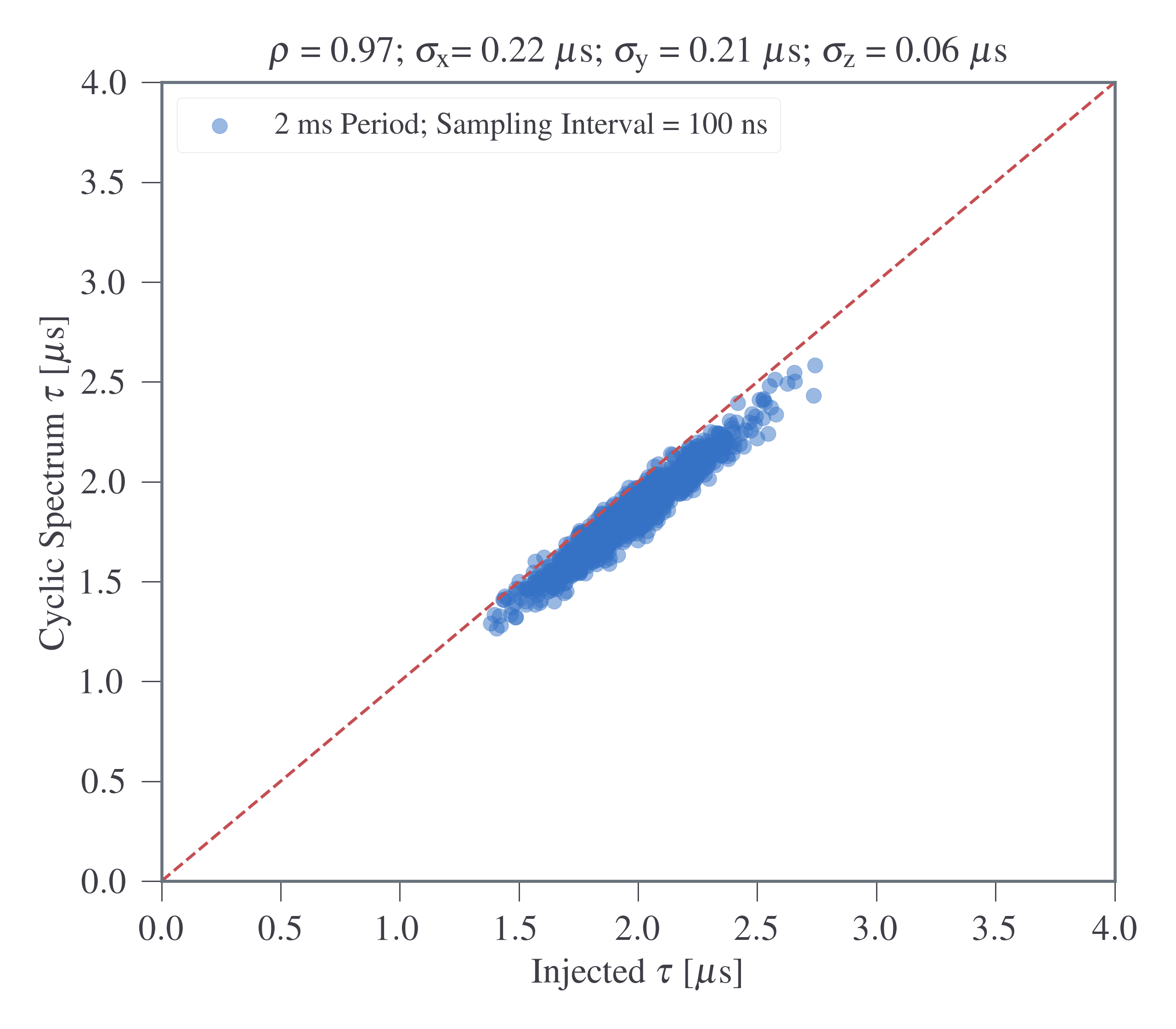}}
\\ 
\subfloat[Lorentzian ACF estimator]{\includegraphics[scale=.54]{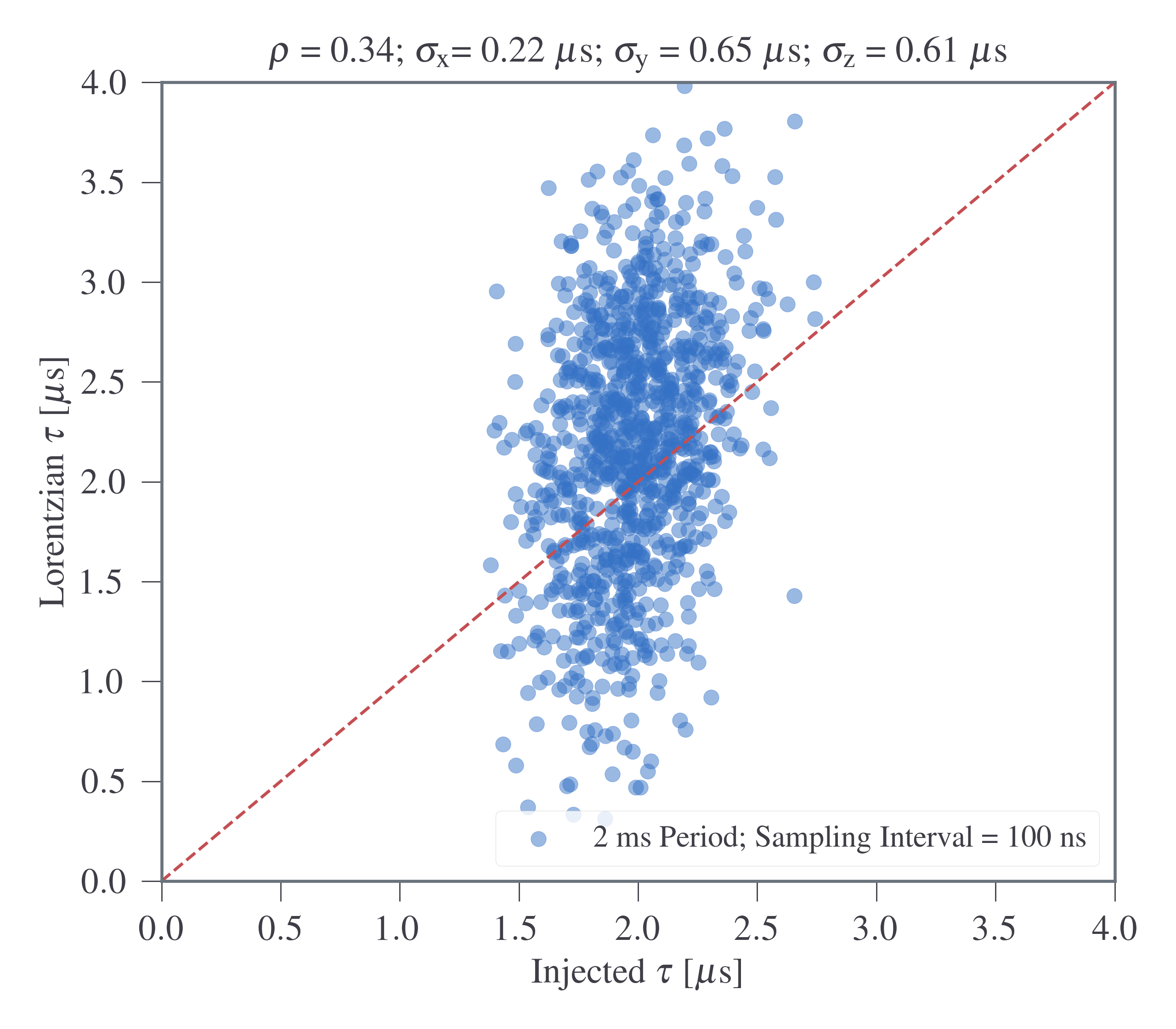}}\hspace{1em}
\subfloat[Gaussian ACF estimator]{\includegraphics[scale=.54]{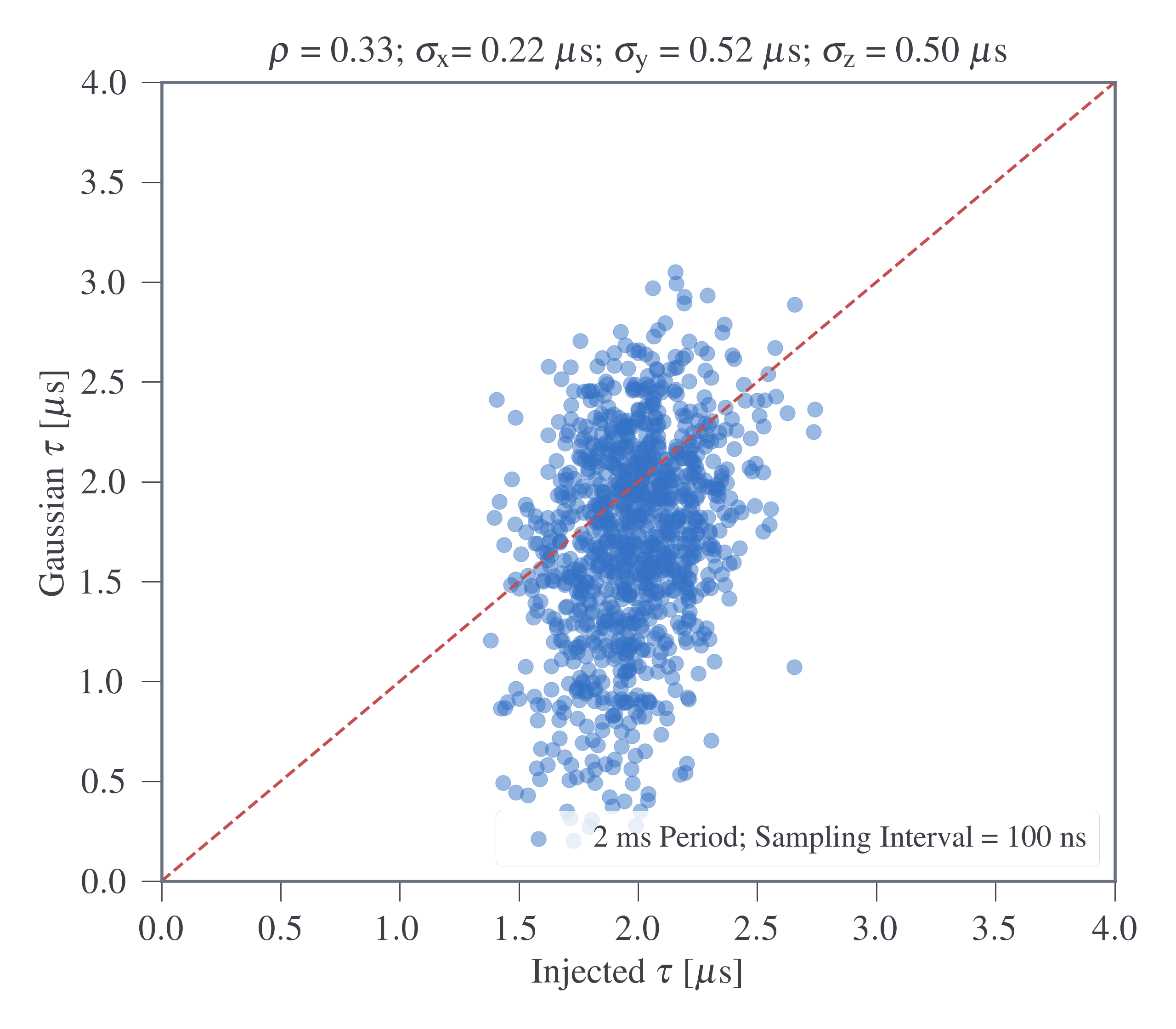}}
\caption{Estimators vs injected delay for 1000 random noise draws for a $\tau$ of 2 $\mu$s using a spin period of 2 ms and a sampling interval of 100 ns at a S/N of 10. $\sigma_x$ is the standard deviation of the data in the $x$ direction, $\sigma_y$ is the spread of the data in the $y$ direction, and $\sigma_z$ is the standard deviation of $z=y-x$. Dashed red lines represent lines of equality between the two axes.}
\label{scatt_inj}
\end{figure*}

 \par These discrepancies become even clearer when we examine how well individual draws correlate between the injected delay and the various estimators for a given S/N. As shown in Figure \ref{scatt_inj}, while there is nearly a one-to-one correspondence between the injected delay and the cyclic spectrum estimator, epoch-to-epoch variations for the ACF estimators are both significantly larger. In these plots $\rho$ represents the correlation coefficient between the two variables, $\sigma_x$ is the standard deviation of the data in the $x$ direction, $\sigma_y$ is the spread of the data in the $y$ direction, and $\sigma_z$ is the standard deviation of $z=y-x$. Additionally, the $\sigma_y$ and $\sigma_z$ values for the ACF estimators are much more similar to each other than to the cyclic spectrum estimator. Larger $\sigma_z$ indicates a larger typical difference between the injected delay and the estimator for a given noise realization.

\par The lack of correlation seen in the ACF estimator plots in Figure \ref{scatt_inj} also present a strong argument against the ACF estimator biases seen in Figure \ref{acf_scatt} simply being an indication that a different $C_1$ should be used for our choice of impulse response, as just choosing a $C_1$ that removes the bias in Figure \ref{acf_scatt} would not alter the lack of correlation seen in Figure \ref{scatt_inj}. The $C_1$ would also have to be different for each ACF approach, since the biases are in opposite directions relative to the injected delay.
 Additionally, attempting to retroactively find $C_1$ by comparing the ratios of the injected delays and ACF-recovered delays in Figure \ref{scatt_inj} shows significant variation among individual realizations in a recovered purported $C_1$.

\par
\begin{figure}[!ht]
\includegraphics[scale=.56]{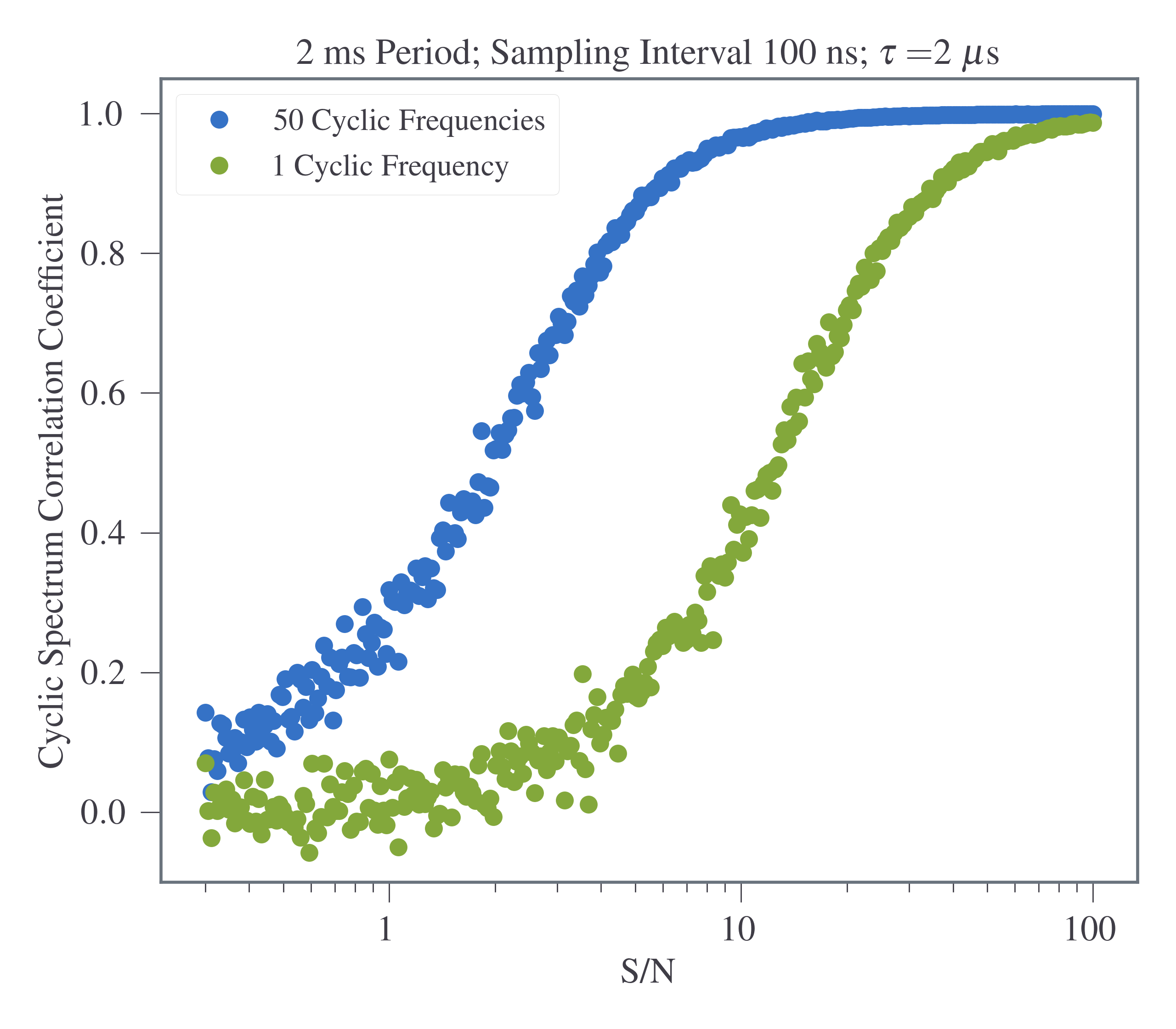}
\centering
\caption{Cyclic spectrum estimator correlation coefficients for random noise draws for a $\tau$ of 2 $\mu$s for 300 values of S/N using a spin period of 2 ms and a sampling interval of 100 ns.}
\label{cyc_corr}
\end{figure}

\begin{figure}[!ht]
\includegraphics[scale=.58]{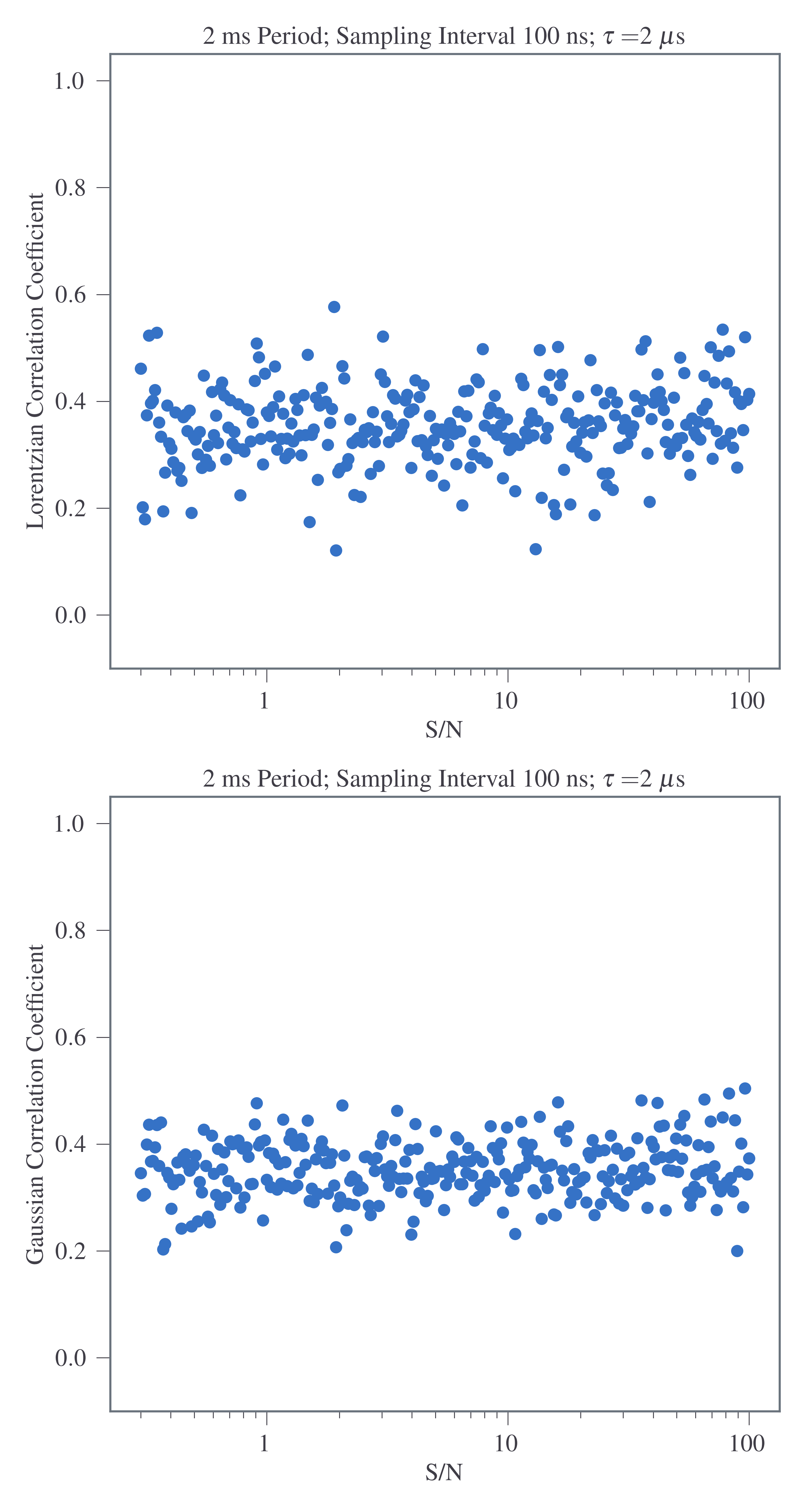}
\centering
\caption{Lorentzian (top) and Gaussian (bottom) ACF estimator correlation coefficients for random noise draws for a $\tau$ of 2 $\mu$s for 300 values of S/N using a spin period of 2 ms and a sampling interval of 100 ns.}
\label{acf_corr}
\end{figure}

 We can also compare how these correlation coefficients change as a function of S/N. For each S/N value, we calculated the correlation coefficients for each estimator over the 1000 random draws. The results are shown in Figures \ref{cyc_corr} and \ref{acf_corr}. As with Figures \ref{cyc_scatt} and \ref{acf_scatt}, we see the cyclic spectrum estimator eventually converge whereas the ACF estimators have already converged. The convergence in the ACF plots, like in Figure \ref{acf_scatt}, are the result of already having sufficient frequency resolution and a sufficient number of scintles over our S/N range, as once the scintle structure in the dynamic spectrum has been resolved, further improvements in S/N will not affect an ACF estimator's ability to recover scattering delays. For the cyclic spectrum estimator, the S/N where it plateaus corresponds well with what is seen in Figure \ref{cyc_scatt}.  Significantly, the cyclic spectrum correlation plateaus at a much higher value than the ACF estimators (around 1.0 compared to around 0.25$-$0.45). This behavior further indicates the improvement the cyclic spectrum estimator provides over the ACF estimators. Additionally, our 50 cyclic frequency estimator converges at a S/N around an order of magnitude earlier than the single cyclic frequency estimator, further emphasizing the benefits of utilizing multiple cyclic frequencies.

\par
\begin{figure}[!ht]
\includegraphics[scale=.58]{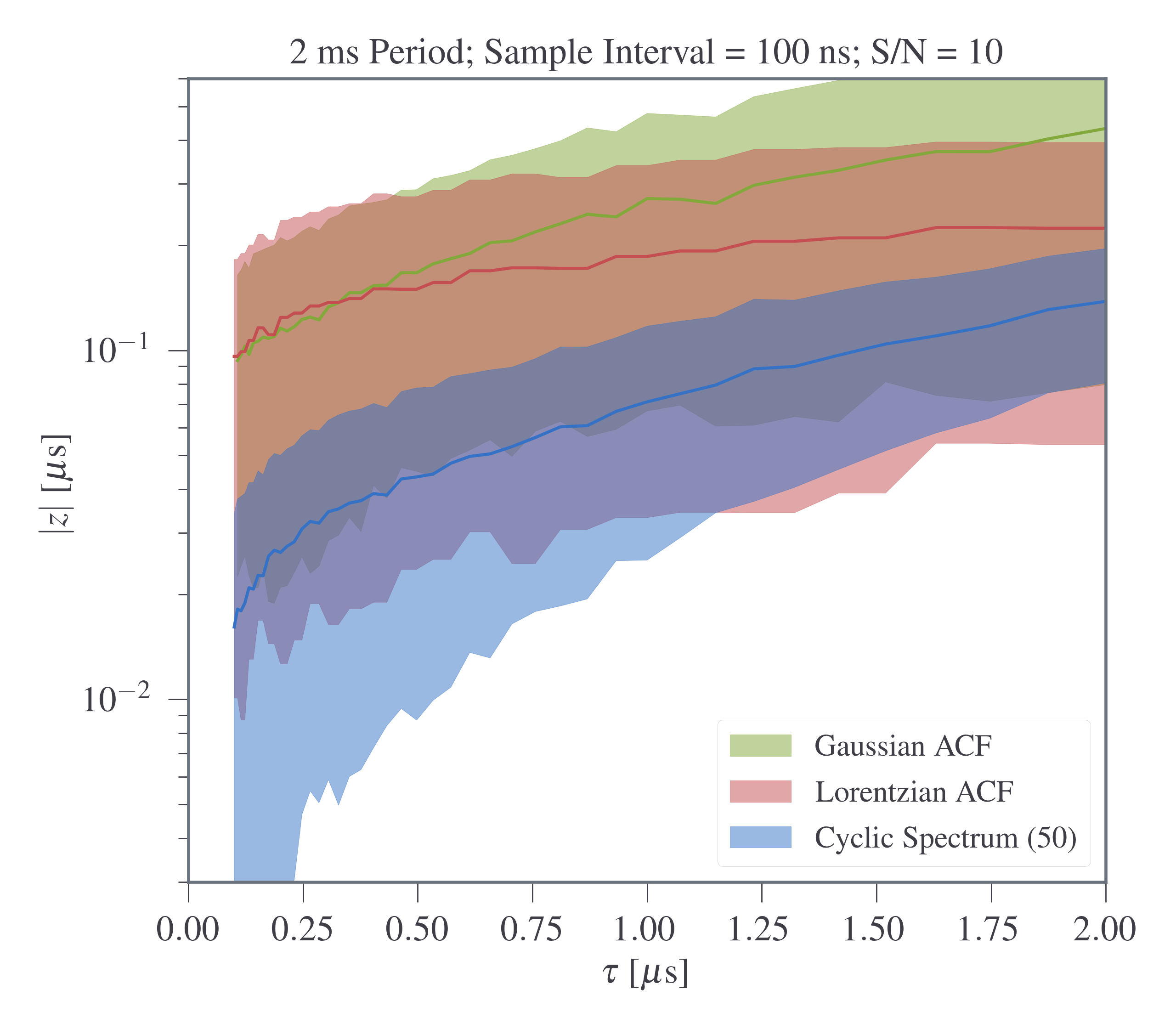}
\centering
\caption{Absolute differences, $|z|$, between the various estimators and $\tau_{\textrm{cent}}$ as a function of the average injected scattering delay, $\tau$. For each of the three curves, the solid line indicates the mean value and the shaded region indicates the 1-$\sigma$ error range.}
\label{tau_z}
\end{figure}

 We also examined how much these estimators deviate from $\tau_{\textrm{cent}}$ as we vary the injected scattering delay. To do this, we repeated the simulation described at the beginning of this section for 45 scattering delays ranging from 0.1$-$2 $\mu$s spaced apart evenly in log space at a S/N of 10. The delay range was chosen based on the breadth of delays we might expect to see from observing many PTA-quality pulsars. We then compared $|z|$, the differences between the estimators and $\tau_{\textrm{cent}}$, over each delay in that range.  The results are shown in Figure \ref{tau_z}. While at the lowest delays for this sampling interval the Gaussian estimator has greater accuracy than the Lorentzian estimator, at higher delays both the Lorentzian and cyclic spectrum estimators are noticeably more accurate than the Gaussian estimator, which is shown to deviate from $\tau_{\textrm{cent}}$ more significantly as the injected delay increases. 

\section{Conclusions and Future Developments}
\label{sec:future}
We simulated scattering delays from the ISM to test the effectiveness of three delay estimators: fitting Lorentzian and Gaussian distributions to  frequency ACFs calculated from pulsar dynamic spectra to recover the scintillation bandwidth and the cyclic spectrum-derived quantity $\tau_{{\rm CS}}$. We find that, at sufficient S/N, in terms of both precision and accuracy, the cyclic spectrum estimator is superior to both ACF estimators, which are accurate over many realizations, but not  as reliable as the cyclic spectrum estimator on an epoch-to-epoch basis. Importantly, for actual pulsar timing with additional sources of timing noise, ACF and CS estimators are necessary to discriminate between ISM-based propagation delays and other sources of delay. We believe the results described in this paper provide significant motivation for further pursuing CS implementation in general, especially through the lens of deconvolution-based IRF recovery.

\par As PTAs close in on sensitivities sufficient for detecting gravitational waves, understanding and mitigating all non-gravitational wave delays will be critical for accurate gravitational wave characterization. Many pulsars in the NANOGrav PTA are already known to have scattering delays of 10s of nanoseconds, which is a non-negligible fraction of the $\mu$s to sub-$\mu$s residuals we see in many pulsars \citep{nanotime}. Many of these estimations, and indeed many estimations of scattering delays in millisecond pulsars, have been performed by fitting Gaussian functions to ACFs, indicating the true effects of scattering delays in PTAs may currently be improperly estimated by a few percent, although additional efforts within NANOGrav are currently in place to estimate scattering delays by fitting $\nu^{-4}$ delays to output TOAs. Additionally, these effects are not currently accounted for in NANOGrav's timing pipeline, or the pipelines of other PTAs such as the European Pulsar Timing Array (EPTA), Parkes Pulsar Timing Array (PPTA), or, consequently, the global pulsar timing array effort, the International Pulsar Timing Array (IPTA), furthering the need for more accurate techniques such as cyclic spectroscopy to be both developed and implemented into future pulsar timing efforts. Efforts are currently ongoing to implement a real-time cyclic spectroscopy backend into existing timing pipelines with the goal of removing scattering effects before any further timing analysis has taken place. This work is currently being done on pipelines operating at the Green Bank Telescope, currently the primary observing site for NANOGrav, but may be implemented in the future at other NANOGrav telescopes such as CHIME and VLA or next-generation telescopes such as DSA-2000 should this endeavor prove successful. 

\vskip 0.3in
This material is based upon work supported by the Green Bank Observatory which is a major facility funded by the National Science Foundation operated by Associated Universities, Inc.
We gratefully acknowledge support of this effort from the NSF Physics Frontiers Center grants 1430284 and 2020265 to NANOGrav. DRS acknowledges support from NSF grant 2009759. TD is supported by an NSF Astronomy and Astrophysics Grant (AAG) award number 2009468. Some of the simulations in this work utilized the resources of the Bowser \& Link computing clusters at West Virginia University.


\par \textit{Software}: \textsc{scipy} \cite{scipy}, \textsc{numpy} \cite{numpy}, and \textsc{matplotlib} \cite{matplotlib}.

\pagebreak

\appendix

\section{Derivation of Fractional Error and Uncertainty As Related To Observable Quantities}

\par The following work uses Equations A9 and A10 of Appendix A of \cite{dsj+20}. The first part of A9 defines cyclic merit as 
\begin{equation}
    m_{\textrm{cyc}} = \frac{b}{\delta b}, 
\end{equation}
where $b=2\pi\tau_{{\rm CS}}/P$ and $\delta b = 2\pi\delta\tau_{{\rm CS}}/P$, while equation A10 defines it as 

\begin{equation}
    m_{\textrm{cyc}} = \frac{2\pi \tau_{{\rm CS}} W_e}{P^2}\textrm{(S/N)}\sqrt{\sum_k k^2 a_k},
\end{equation}
where $W_e$ is the effective pulse width and $a_k=A_k/A_0$, the ratio of the k$^{\textrm{th}}$ coefficient and the 0$^{\textrm{th}}$ coefficient of the intensity pulse profile's Fourier transform. For a sharp pulse, Fourier coefficients should stay substantial out to some high number $k_{\textrm{max}}$ before falling off rapidly, with the number of cyclic frequencies we go up to being roughly the inverse of the duty cycle. This means that $k_{\textrm{max}}$ should be roughly $P/W_e$. If we assume that $a_k$ stays constant out to $k_\textrm{max}$, the radical becomes 

\begin{equation}
\sqrt{\sum_{k=1}^{k_{\textrm{max}}} k^2} = \sqrt{\frac{k_{\textrm{max}}(k_{\textrm{max}}+1)(2k_{\textrm{max}}+1)}{6}}\approx \sqrt{k_{\textrm{max}}^3}\approx \Bigg(\frac{P}{W_e}\Bigg)^{3/2}.
\end{equation}

From here we can say that 
\begin{equation}
    \frac{b}{\delta b} \approx \frac{2\pi \tau_{{\rm CS}} W_e}{P^2}\textrm{(S/N)}\Bigg(\frac{P}{W_e}\Bigg)^{3/2} = \frac{2\pi\tau_{{\rm CS}} \textrm{(S/N)}}{\sqrt{P W_e}}.
\end{equation}
We can then express this inverse fractional error as
\begin{align}
    \frac{b}{\delta b}&= \frac{2\pi \tau_{{\rm CS}}}{P \delta b} =\frac{2\pi\tau_{{\rm CS}} \textrm{(S/N)}}{\sqrt{P W_e}} \\ \implies \delta b & = \sqrt{\frac{W_e}{P}}\frac{1}{\textrm{(S/N)}},
\end{align}
meaning the uncertainty in $\tau_{{\rm CS}}$ can be expressed as

\begin{equation}
    \delta \tau_{{\rm CS}} = \frac{\sqrt{P W_e}}{2\pi \textrm{(S/N)}}.
\end{equation}

\bibliographystyle{Refs/apj} 
\bibliography{Refs/journals_apj,Refs/scintillation,Refs/modrefs,Refs/modrefs_drs, Refs/psrrefs}

\end{document}

%% file: authors.tex
\author[0000-0002-2451-7288]{Jacob E. Turner}
\affiliation{Department of Physics and Astronomy, West Virginia University, P.O. Box 6315, Morgantown, WV 26506, USA}
\affiliation{Center for Gravitational Waves and Cosmology, West Virginia University, Chestnut Ridge Research Building, Morgantown, WV 26505, USA}

\author[0000-0002-1797-3277]{Daniel R. Stinebring}
\affiliation{Department of Physics and Astronomy, Oberlin College, Oberlin, OH 44074, USA}

\author[0000-0001-7697-7422]{Maura A. McLaughlin}
\affiliation{Department of Physics and Astronomy, West Virginia University, P.O. Box 6315, Morgantown, WV 26506, USA}
\affiliation{Center for Gravitational Waves and Cosmology, West Virginia University, Chestnut Ridge Research Building, Morgantown, WV 26505, USA}
\author[0000-0003-0638-3340]{Anne M. Archibald}
\affiliation{Newcastle University, Newcastle upon Tyne NE1 7RU UK}
\author[0000-0001-8885-6388]{Timothy Dolch}
\affiliation{Department of Physics, Hillsdale College, 33 E. College Street, Hillsdale, MI 49242, USA}
\affiliation{Eureka Scientific, 2452 Delmer Street, Suite 100, Oakland, CA 94602-3017, USA }
\author[0000-0001-5229-7430]{Ryan S. Lynch}
\affiliation{Green Bank Observatory, P.O. Box 2, Green Bank, WV 24944, USA}

%% file: abstract.tex
\begin{abstract}
We simulate scattering delays from the interstellar medium to examine the effectiveness of three estimators in recovering these delays in pulsar timing data. Two of these estimators use the more traditional process of fitting autocorrelation functions to pulsar dynamic spectra to extract scintillation bandwidths, while the third estimator uses the newer technique of cyclic spectroscopy on baseband pulsar data to recover the interstellar medium's impulse response function. We find that either fitting a Lorentzian or Gaussian distribution to an autocorrelation function or recovering the impulse response function from the cyclic spectrum are, on average, accurate in recovering scattering delays, although autocorrelation function estimators have a large variance, even at high signal-to-noise ratio (S/N). We find that, given sufficient S/N, cyclic spectroscopy is more accurate than both Gaussian and Lorentzian fitting for recovering scattering delays at specific epochs, suggesting that cyclic spectroscopy is a superior method for scattering estimation in high quality data.
\end{abstract}

%% file: Turner_estimating_tau_scatt.bbl
\begin{thebibliography}{}
\expandafter\ifx\csname natexlab\endcsname\relax\def\natexlab#1{#1}\fi

\bibitem[{Alam {et~al.}(2020)Alam, Arzoumanian, Baker, Blumer, Bohler, Brazier,
  Brook, Burke-Spolaor, Caballero, Camuccio, Chamberlain, Chatterjee, Cordes,
  Cornish, Crawford, Cromartie, DeCesar, Demorest, Dolch, Ellis, Ferdman,
  Ferrara, Fiore, Fonseca, Garcia, Garver-Daniels, Gentile, Good, Gusdorff,
  Halmrast, Hazboun, Islo, Jennings, Jessup, Jones, Kaiser, Kaplan, Kelley,
  Key, Lam, Lazio, Lorimer, Luo, Lynch, Madison, Maraccini, McLaughlin,
  Mingarelli, Ng, Nguyen, Nice, Pennucci, Pol, Ramette, Ransom, Ray,
  Shapiro-Albert, Siemens, Simon, Spiewak, Stairs, Stinebring, Stovall,
  Swiggum, Taylor, Tripepi, Vallisneri, Vigeland, Witt, Zhu, \&
  Collaboration)}]{nanotime}
Alam, M.~F., Arzoumanian, Z., Baker, P.~T., {et~al.} 2020, The Astrophysical
  Journal Supplement Series, 252, 4

\bibitem[{Antoni(2007)}]{ANTONI2007597}
Antoni, J. 2007, Mechanical Systems and Signal Processing, 21, 597

\bibitem[{{Antoniadis} {et~al.}(2022){Antoniadis}, {Arzoumanian}, {Babak},
  {Bailes}, {Bak Nielsen}, {Baker}, {Bassa}, {B{\'e}csy}, {Berthereau},
  {Bonetti}, {Brazier}, {Brook}, {Burgay}, {Burke-Spolaor}, {Caballero},
  {Casey-Clyde}, {Chalumeau}, {Champion}, {Charisi}, {Chatterjee}, {Chen},
  {Cognard}, {Cordes}, {Cornish}, {Crawford}, {Cromartie}, {Crowter}, {Dai},
  {DeCesar}, {Demorest}, {Desvignes}, {Dolch}, {Drachler}, {Falxa}, {Ferrara},
  {Fiore}, {Fonseca}, {Gair}, {Garver-Daniels}, {Goncharov}, {Good}, {Graikou},
  {Guillemot}, {Guo}, {Hazboun}, {Hobbs}, {Hu}, {Islo}, {Janssen}, {Jennings},
  {Johnson}, {Jones}, {Kaiser}, {Kaplan}, {Karuppusamy}, {Keith}, {Kelley},
  {Kerr}, {Key}, {Kramer}, {Lam}, {Lamb}, {Lazio}, {Lee}, {Lentati}, {Liu},
  {Luo}, {Lynch}, {Lyne}, {Madison}, {Main}, {Manchester}, {McEwen}, {McKee},
  {McLaughlin}, {Mickaliger}, {Mingarelli}, {Ng}, {Nice}, {Os{\l}owski},
  {Parthasarathy}, {Pennucci}, {Perera}, {Perrodin}, {Petiteau}, {Pol},
  {Porayko}, {Possenti}, {Ransom}, {Ray}, {Reardon}, {Russell}, {Samajdar},
  {Sampson}, {Sanidas}, {Sarkissian}, {Schmitz}, {Schult}, {Sesana},
  {Shaifullah}, {Shannon}, {Shapiro-Albert}, {Siemens}, {Simon}, {Smith},
  {Speri}, {Spiewak}, {Stairs}, {Stappers}, {Stinebring}, {Swiggum}, {Taylor},
  {Theureau}, {Tiburzi}, {Vallisneri}, {van der Wateren}, {Vecchio},
  {Verbiest}, {Vigeland}, {Wahl}, {Wang}, {Wang}, {Wang}, {Witt}, {Zhang}, \&
  {Zhu}}]{2022MNRAS.510.4873A}
{Antoniadis}, J., {Arzoumanian}, Z., {Babak}, S., {et~al.} 2022, \mnras, 510,
  4873

\bibitem[{Archibald {et~al.}(2014)Archibald, Kondratiev, Hessels, \&
  Stinebring}]{Archibald_2014}
Archibald, A.~M., Kondratiev, V.~I., Hessels, J. W.~T., \& Stinebring, D.~R.
  2014, The Astrophysical Journal, 790, L22

\bibitem[{{Arzoumanian} {et~al.}(2020){Arzoumanian}, {Baker}, {Blumer},
  {B{\'e}csy}, {Brazier}, {Brook}, {Burke-Spolaor}, {Chatterjee}, {Chen},
  {Cordes}, {Cornish}, {Crawford}, {Cromartie}, {Decesar}, {Demorest}, {Dolch},
  {Ellis}, {Ferrara}, {Fiore}, {Fonseca}, {Garver-Daniels}, {Gentile}, {Good},
  {Hazboun}, {Holgado}, {Islo}, {Jennings}, {Jones}, {Kaiser}, {Kaplan},
  {Kelley}, {Key}, {Laal}, {Lam}, {Lazio}, {Lorimer}, {Luo}, {Lynch},
  {Madison}, {McLaughlin}, {Mingarelli}, {Ng}, {Nice}, {Pennucci}, {Pol},
  {Ransom}, {Ray}, {Shapiro-Albert}, {Siemens}, {Simon}, {Spiewak}, {Stairs},
  {Stinebring}, {Stovall}, {Sun}, {Swiggum}, {Taylor}, {Turner}, {Vallisneri},
  {Vigeland}, {Witt}, \& {Nanograv Collaboration}}]{nano_12}
{Arzoumanian}, Z., {Baker}, P.~T., {Blumer}, H., {et~al.} 2020, \apjl, 905, L34

\bibitem[{Bansal {et~al.}(2019)Bansal, Taylor, Stovall, \&
  Dowell}]{Bansal_2019}
Bansal, K., Taylor, G.~B., Stovall, K., \& Dowell, J. 2019, The Astrophysical
  Journal, 875, 146

\bibitem[{{Bhat} {et~al.}(2004){Bhat}, {Cordes}, {Camilo}, {Nice}, \&
  {Lorimer}}]{bcc+04}
{Bhat}, N.~D.~R., {Cordes}, J.~M., {Camilo}, F., {Nice}, D.~J., \& {Lorimer},
  D.~R. 2004, ApJ, 605, 759

\bibitem[{Bhat {et~al.}(1999)Bhat, Rao, \& Gupta}]{Bhat_1999}
Bhat, N. D.~R., Rao, A.~P., \& Gupta, Y. 1999, The Astrophysical Journal
  Supplement Series, 121, 483

\bibitem[{Brown \& Loomis(1993)}]{cyc3}
Brown, W., \& Loomis, H. 1993, IEEE Transactions on Signal Processing, 41, 703

\bibitem[{Chen {et~al.}(2021)Chen, Caballero, Guo, Chalumeau, Liu, Shaifullah,
  Lee, Babak, Desvignes, Parthasarathy, Hu, van der Wateren, Antoniadis,
  Bak Nielsen, Bassa, Berthereau, Burgay, Champion, Cognard, Falxa, Ferdman,
  Freire, Gair, Graikou, Guillemot, Jang, Janssen, Karuppusamy, Keith, Kramer,
  Liu, Lyne, Main, McKee, Mickaliger, Perera, Perrodin, Petiteau, Porayko,
  Possenti, Samajdar, Sanidas, Sesana, Speri, Stappers, Theureau, Tiburzi,
  Vecchio, Verbiest, Wang, Wang, \& Xu}]{epta_back}
Chen, S., Caballero, R.~N., Guo, Y.~J., {et~al.} 2021, Monthly Notices of the
  Royal Astronomical Society, 508, 4970

\bibitem[{{Coles} {et~al.}(2010){Coles}, {Rickett}, {Gao}, {Hobbs}, \&
  {Verbiest}}]{coles_2010}
{Coles}, W.~A., {Rickett}, B.~J., {Gao}, J.~J., {Hobbs}, G., \& {Verbiest},
  J.~P.~W. 2010, ApJ, 717, 1206

\bibitem[{Cordes(1986)}]{cor86}
Cordes, J.~M. 1986, ApJ, 311, 183

\bibitem[{{Cordes} \& {Rickett}(1998)}]{cr98}
{Cordes}, J.~M., \& {Rickett}, B.~J. 1998, ApJ, 507, 846

\bibitem[{{Cordes} {et~al.}(2016){Cordes}, {Shannon}, \& {Stinebring}}]{css16}
{Cordes}, J.~M., {Shannon}, R.~M., \& {Stinebring}, D.~R. 2016, \apj, 817, 16

\bibitem[{Cordes {et~al.}(1985)Cordes, Weisberg, \& Boriakoff}]{cwb85}
Cordes, J.~M., Weisberg, J.~M., \& Boriakoff, V. 1985, ApJ, 288, 221

\bibitem[{{Demorest}(2011)}]{dem11}
{Demorest}, P.~B. 2011, MNRAS, 416, 2821

\bibitem[{Dolch {et~al.}(2021)Dolch, Stinebring, Jones, Zhu, Lynch, Cohen,
  Demorest, Lam, Levin, McLaughlin, \& Palliyaguru}]{dsj+20}
Dolch, T., Stinebring, D.~R., Jones, G., {et~al.} 2021, The Astrophysical
  Journal, 913, 98

\bibitem[{Gardner(1987)}]{cyc1}
Gardner, W. 1987, IEEE Transactions on Communications, 35, 584

\bibitem[{Goncharov {et~al.}(2021)Goncharov, Shannon, Reardon, Hobbs, Zic,
  Bailes, Curyło, Dai, Kerr, Lower, Manchester, Mandow, Middleton, Miles,
  Parthasarathy, Thrane, Thyagarajan, Xue, Zhu, Cameron, Feng, Luo, Russell,
  Sarkissian, Spiewak, Wang, Wang, Zhang, \& Zhang}]{ppta_back}
Goncharov, B., Shannon, R.~M., Reardon, D.~J., {et~al.} 2021, The Astrophysical
  Journal Letters, 917, L19

\bibitem[{Hemberger \& Stinebring(2008)}]{hs08}
Hemberger, D.~A., \& Stinebring, D.~R. 2008, ApJ, 674, L37

\bibitem[{{Hill} {et~al.}(2005){Hill}, {Stinebring}, {Asplund}, {Berwick},
  {Everett}, \& {Hinkel}}]{hsa+05}
{Hill}, A.~S., {Stinebring}, D.~R., {Asplund}, C.~T., {et~al.} 2005, ApJL, 619,
  L171

\bibitem[{Hunter(2007)}]{matplotlib}
Hunter, J.~D. 2007, Computing in Science \& Engineering, 9, 90

\bibitem[{{Jones} {et~al.}(2017){Jones}, {McLaughlin}, {Lam}, {Cordes},
  {Levin}, {Chatterjee}, {Arzoumanian}, {Crowter}, {Demorest}, {Dolch},
  {Ellis}, {Ferdman}, {Fonseca}, {Gonzalez}, {Jones}, {Lazio}, {Nice},
  {Pennucci}, {Ransom}, {Stinebring}, {Stairs}, {Stovall}, {Swiggum}, \&
  {Zhu}}]{jones_dm}
{Jones}, M.~L., {McLaughlin}, M.~A., {Lam}, M.~T., {et~al.} 2017, \apj, 841,
  125

\bibitem[{{Levin} {et~al.}(2016){Levin}, {McLaughlin}, {Jones}, {Cordes},
  {Stinebring}, {Chatterjee}, {Dolch}, {Lam}, {Lazio}, {Palliyaguru},
  {Arzoumanian}, {Crowter}, {Demorest}, {Ellis}, {Ferdman}, {Fonseca},
  {Gonzalez}, {Jones}, {Nice}, {Pennucci}, {Ransom}, {Stairs}, {Stovall},
  {Swiggum}, \& {Zhu}}]{levin_scat}
{Levin}, L., {McLaughlin}, M.~A., {Jones}, G., {et~al.} 2016, \apj, 818, 166

\bibitem[{Main {et~al.}(2017)Main, van Kerkwijk, Pen, Mahajan, \&
  Vanderlinde}]{Main_2017}
Main, R., van Kerkwijk, M., Pen, U.-L., Mahajan, N., \& Vanderlinde, K. 2017,
  The Astrophysical Journal Letters, 840, L15

\bibitem[{{Manchester} {et~al.}(2013){Manchester}, {Hobbs}, {Bailes}, {Coles},
  {van Straten}, {Keith}, {Shannon}, {Bhat}, {Brown}, {Burke-Spolaor},
  {Champion}, {Chaudhary}, {Edwards}, {Hampson}, {Hotan}, {Jameson}, {Jenet},
  {Kesteven}, {Khoo}, {Kocz}, {Maciesiak}, {Oslowski}, {Ravi}, {Reynolds},
  {Sarkissian}, {Verbiest}, {Wen}, {Wilson}, {Yardley}, {Yan}, \&
  {You}}]{parkes}
{Manchester}, R.~N., {Hobbs}, G., {Bailes}, M., {et~al.} 2013, \pasa, 30, e017

\bibitem[{Narayan \& Goodman(1989)}]{Narayan}
Narayan, R., \& Goodman, J. 1989, Monthly Notices of the Royal Astronomical
  Society, 238, 963

\bibitem[{Palliyaguru {et~al.}(2015)Palliyaguru, Stinebring, McLaughlin,
  Demorest, \& Jones}]{Palliyaguru_2015}
Palliyaguru, N., Stinebring, D., McLaughlin, M., Demorest, P., \& Jones, G.
  2015, The Astrophysical Journal, 815, 89

\bibitem[{{Rickett}(1975)}]{ric75}
{Rickett}, B.~J. 1975, \apj, 197, 185

\bibitem[{Roberts {et~al.}(1991)Roberts, Brown, \& Loomis}]{cyc2}
Roberts, R., Brown, W., \& Loomis, H. 1991, IEEE Signal Processing Magazine, 8,
  38

\bibitem[{{Turner} {et~al.}(2021){Turner}, {McLaughlin}, {Cordes}, {Lam},
  {Shapiro-Albert}, {Stinebring}, {Arzoumanian}, {Blumer}, {Brook},
  {Chatterjee}, {Cromartie}, {DeCesar}, {Demorest}, {Dolch}, {Ellis},
  {Ferdman}, {Ferrara}, {Fonseca}, {Garver-Daniels}, {Gentile}, {Good},
  {Jones}, {Lazio}, {Lorimer}, {Luo}, {Lynch}, {Ng}, {Nice}, {Pennucci}, {Pol},
  {Ransom}, {Spiewak}, {Stairs}, {Stovall}, {Swiggum}, \&
  {Vigeland}}]{turner_scat}
{Turner}, J.~E., {McLaughlin}, M.~A., {Cordes}, J.~M., {et~al.} 2021, \apj,
  917, 10

\bibitem[{{van der Walt} {et~al.}(2011){van der Walt}, {Colbert}, \&
  {Varoquaux}}]{numpy}
{van der Walt}, S., {Colbert}, S.~C., \& {Varoquaux}, G. 2011, Computing in
  Science Engineering, 13, 22

\bibitem[{{Virtanen} {et~al.}(2020){Virtanen}, {Gommers}, {Oliphant},
  {Haberland}, {Reddy}, {Cournapeau}, {Burovski}, {Peterson}, {Weckesser},
  {Bright}, {van der Walt}, {Brett}, {Wilson}, {Jarrod Millman}, {Mayorov},
  {Nelson}, {Jones}, {Kern}, {Larson}, {Carey}, {Polat}, {Feng}, {Moore}, {Vand
  erPlas}, {Laxalde}, {Perktold}, {Cimrman}, {Henriksen}, {Quintero}, {Harris},
  {Archibald}, {Ribeiro}, {Pedregosa}, {van Mulbregt}, \&
  {Contributors}}]{scipy}
{Virtanen}, P., {Gommers}, R., {Oliphant}, T.~E., {et~al.} 2020, Nature
  Methods, 17, 261

\bibitem[{Walker {et~al.}(2013)Walker, Demorest, \& van Straten}]{Walker_2013}
Walker, M.~A., Demorest, P.~B., \& van Straten, W. 2013, The Astrophysical
  Journal, 779, 99

\bibitem[{{Walker} {et~al.}(2008){Walker}, {Koopmans}, {Stinebring}, \& {van
  Straten}}]{wksv08}
{Walker}, M.~A., {Koopmans}, L.~V.~E., {Stinebring}, D.~R., \& {van Straten},
  W. 2008, MNRAS, 388, 1214

\bibitem[{{Wang} {et~al.}(2005){Wang}, {Manchester}, {Johnston}, {Rickett},
  {Zhang}, {Yusup}, \& {Chen}}]{wmj+05}
{Wang}, N., {Manchester}, R.~N., {Johnston}, S., {et~al.} 2005, MNRAS, 358, 270

\end{thebibliography}
